\documentclass[aps,prb,amsmath,amssymb,reprint]{revtex4-2}
\usepackage{graphicx}
\usepackage{epstopdf}
\usepackage{amsmath}
\usepackage{indentfirst}
\usepackage{float}
\usepackage{subfigure}

\usepackage[utf8]{inputenc}
\usepackage[T1]{fontenc}

\begin{document}

\title{Quantum Resistor-Capacitor Circuit with two Majorana Bound States}

\author{Cong Li}
\affiliation{Key Laboratory of Artificial Structures and Quantum Control (Ministry of Education), Department of Physics and 
Astronomy, Shanghai Jiaotong University, 800 Dongchuan Road, Shanghai 200240, China}

\author{Bing Dong}
\thanks{Author to whom correspondence should be addressed. Email:bdong@sjtu.edu.cn.}
\affiliation{Key Laboratory of Artificial Structures and Quantum Control (Ministry of Education), Department of Physics and 
Astronomy, Shanghai Jiaotong University, 800 Dongchuan Road, Shanghai 200240, China}

\begin{abstract}

In this paper, we derive the equations of motion for a system composed of a spinless quantum dot coupled to two normal leads and two Majorana bound states (MBSs), utilizing the auxiliary-mode expansion method and the nonequilibrium Green function technique. Subsequently, we use these equations to analyze the linear conductance, adiabatic linear capacitance, and adiabatic linear relaxation resistance of the system.
We find that when the phase difference between the two MBSs is an integer multiple of
$\pi$, the MBSs enter the zero mode, leading to the complete suppression of the linear relaxation resistance. In this case, the relaxation resistance is highly sensitive to the MBSs modes. On the other hand, when the phase difference is not an integer multiple of
$\pi$, the linear conductance is fully suppressed. Furthermore, the linear relaxation resistance remains completely suppressed, even when the MBSs are not in the zero mode, and the system loses its sensitivity to the MBSs modes.
\end{abstract}
\maketitle

\section{Introduction}

Majorana bound states (MBSs) have emerged as a focal point of intense research in both experimental\cite{RN24,RN25,RN105,RN108,RN131,RN132,RN134,RN135,RN136} and theoretical studies\cite{RN40,RN56,RN41,RN74,RN52,RN45,RN46,RN32,RN72,RN66,RN53,RN42,RN87,RN28}, owing to their profound fundamental implications and promising applications in quantum computation\cite{RN200,RN201,RN202}. In mesoscopic systems, such as a quantum dot (QD) coupled to two normal leads and an MBS, these states give rise to distinctive electrical transport phenomena. Notably, they can induce a zero-bias conductance peak corresponding to half of the unitary conductance ($G_{0} = e^2/h$)\cite{RN27,RN28}. Furthermore, when a QD is coupled to multiple MBSs, the phase differences between the MBSs play a critical role in modulating the system's transport properties\cite{RN27,RN57,RN80,RN69,RN7,RN107}.

Recent investigations have extended to quantum resistor-capacitor circuits, where a spinless QD is interfaced with an MBS localized at the edges of two-dimensional topological superconductors\cite{RN36} or embedded within topological superconducting wires\cite{RN8}. These studies revealed that when the MBS is in the zero-mode, the system's adiabatic linear relaxation resistance is entirely suppressed.
Their investigation further revealed that the mode of MBS can significantly influence the magnitude of the linear capacitance in the system.
On the other hand, recent years have seen burgeoning interest in systems hosting multiple MBSs\cite{twoMBSs1,twoMBSs2,twoMBSs3,twoMBSs4,twoMBSs5,twoMBSs6,twoMBSs7}, which demonstrate remarkable phenomena such as phase difference induced complete conductance suppression\cite{RN27,twoMBSs1,twoMBSs8,RN7,RN80} and nonlocal effects mediated by terminal MBSs in nanowires\cite{twoMBSs2,twoMBSs3,twoMBSs5,twoMBSs7}. Furthermore, the characteristic modes of MBSs can significantly influence the periodic behavior of the system's conductance as a function of the phase difference between the MBSs\cite{twoMBSs9}. Nevertheless, the investigation of quantum RC circuits incorporating two MBSs remains unexplored territory. Considering the established profound influence of inter-MBS phase differences on conductance characteristics, we posit that this phase parameter would similarly govern the RC response behavior in such systems, presenting a compelling avenue for research.
To date, no pertinent experimental research has been undertaken in this domain, leaving these crucial investigations as important future work to be pursued.

In this study, we investigate both the linear conductance and adiabatic alternating current (AC) response in a hybrid quantum system, which is composed of a spinless QD coupled to two MBSs
of a nanowire
and two normal metallic leads. For this purpose, a suitable theoretical formalism is essential. It should be capable of exactly capturing the transient dynamics of this hybrid quantum system under arbitrary time-dependent external fields. In the literature, the nonequilibrium Green's function (NGF) is widely regarded as the most powerful method for investigation of nonequilibrium time-dependent transport problem in a broad range of nanoscale systems. However, while not impossible, the direct computation of the double-time NGF in this context demands such an extensive amount numerical calculations that it becomes impractical.
To circumvent this challenge, a novel propagation scheme has recently been developed, specifically tailored for transient transport in a noninteracting QD. This scheme commences with the equations of motion (EOMs) of the density matrix and auxiliary current matrices. By utilizing the time-dependent NGF, this approach ultimately yields a set of coupled ordinary differential equations characterized by a single time argument. This is accomplished through the application of an auxiliary-mode expansion technique\cite{RN203,RN204,RN205,RN119,RN127,RN112}. This scheme has been successfully implemented to study electron pumps and time-dependent transport in molecular junctions. One of the authors has further generalized this propagation scheme to accommodate the strong electron correlation effects in time-dependent nonequilibrium transport\cite{RN112}, such as the dynamic Kondo-type tunneling through an interacting QD. In this paper, we first generalize the propagation scheme to be capable of dealing with the hybrid MBS-QD system, and then apply this method to investigate its low-frequency AC characteristics.

The rest of this paper is organized as follows. In Section II, we introduce the time-dependent model Hamiltonian and outline the theoretical method, namely, the auxiliary-mode expansion technique. We begin with the EOMs of density matrix, then define the auxiliary current matrices, and proceed subsequently to expand them in terms of auxiliary modes. This expansion is accomplished by employing a Pad$\acute{e}$ expansion of the Fermi function and the NGF technique under the assumption of a wide band limit. We further derive the EOMs for the auxiliary-mode expansion of the current matrices, ultimately arriving at a closed set of coupled differential equations. To investigate the adiabatic response of the hybrid MBS-QD system, we expand and solve these equations in the limit of low-frequency modulation signals. Utilizing these solutions, we proceed to calculate the linear capacitance and relaxation resistance.
In Section III, we carry out numerical calculations of the linear conductance, capacitance, and relaxation resistance of the system, and discuss these results. Finally, a brief summary is given in Section IV.

\section{Model Hamiltonian and Theoretical Method}

In this paper, we consider a four-terminal hybrid nanodevice consisting of a spinless QD connected to two normal leads and two MBSs of
a nanowire, as illustrated in Fig. 1.
From an experimental perspective, this setup eliminates the need for bending the superconducting wire containing the MBSs, as it only requires coupling both terminal MBSs to the QD while generating a phase difference in their tunneling couplings to the QD through the application of an external magnetic flux.
The model Hamiltonian of the system is given by:
\begin{equation}
\begin{split}
H(t)&=H_{MBS}(t)+H_{QD}(t)\\
&\;\;\;+H_{QD-M}(t)+H_{Lead}(t)+H_T(t),
\end{split}
\end{equation}
with
\begin{equation}
H_{MBS}(t)=i\varepsilon_m(t)\gamma_1\gamma_2,
\end{equation}
\begin{equation}
H_{QD}(t)=\varepsilon_d(t)d^{\dag}d,
\end{equation}
\begin{equation}
\begin{split}
H_{QD-M}(t)&=V_1(t)d^{\dag}\gamma_1+V_1^{*}(t)\gamma_1d\\
&\;\;\;+V_2(t)d^{\dag}\gamma_2+V_2^{*}(t)\gamma_2d,
\end{split}
\end{equation}
\begin{equation}
H_{Lead}(t)=\sum_{\eta k}\varepsilon_{\eta k}(t)c_{\eta k}^{\dag}c_{\eta k},
\end{equation}
\begin{equation}
H_{T}(t)=\sum_{\eta k}V_{\eta k}(t)(c_{\eta k}^{\dag}d+d^{\dag}c_{\eta k}).
\end{equation}
In the above equations, $\gamma_{1}$ and $\gamma_{2}$ denote the creation and annihilation operators for the two MBSs, while $c_{\eta k}^{\dag}$($c_{\eta k}$) and $d^{\dag}$($d$) are the electron creation (annihilation) operators for the $\eta$-th normal lead and the QD, respectively. In the MBSs Hamiltonian $H_{MBS}(t)$, $\varepsilon_m(t)$ represents the mode of the MBSs. In the QD Hamiltonian $H_{QD}(t)$, $\varepsilon_d(t)$ denotes the bare dot level of the QD. The term $H_{QD-M}(t)$ describes the tunneling process between the QD and the two MBSs, with $V_{1}(t)$ and $V_{2}(t)$ serving as the tunneling parameters. The Hamiltonian of the leads, $H_{Lead}(t)$, indexs $\eta=L,R$ for the left and right leads, and $\varepsilon_{\eta k}(t)$ represents the energy level of electrons in the two normal leads. Lastly, $H_{T}(t)$ represents the tunneling process between the QD and the two normal leads, where $V_{\eta k}(t)$ is the pertinent tunneling matrix element.
The corresponding coupling strength is defined as $\Gamma_{\eta}(t)=\sum_{k}\delta(\varepsilon-\varepsilon_{\eta k}^0)(V_{\eta k}(t))^2$,  which is assumed to be independent of energy in the wide band limit. To methodically account for the temporal modulation, we decompose the coupling matrix elements as $V_{\eta k}(t)=u_{\eta}(t)V_{\eta k}^{0}$, where $u_{\eta}(t)$ represents the dimensionless time-dependent control profile and $V_{\eta k}^{0}$ denotes the bare coupling parameters. This factorization enables the separation of temporal dynamics from static coupling properties, yielding a time-independent base coupling strength: $\Gamma_{\eta}^{0}=\sum_{k}\delta(\varepsilon-\varepsilon_{\eta k}^0)(V_{\eta k}^{0})^2$.
Throughout we will use natural units $\hbar$ = $k_B$ = $e$ = 1.

\begin{figure}[H]
\centering
\includegraphics[scale=0.26]{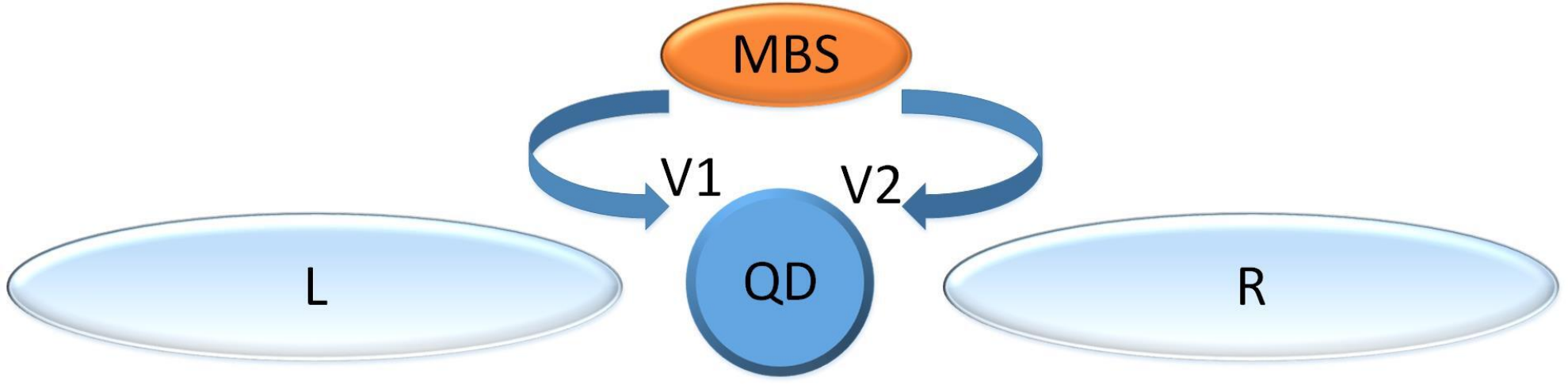}
\caption{(Color online)  Schematic diagram of a spinless QD connected to two normal leads and two MBSs.}
\end{figure}

Using the regular fermion representation: $\gamma_{1}=(f+ f^{\dag})/\sqrt{2}$ and $\gamma_{2}=i(f^{\dag}- f)/\sqrt{2}$, we can rewrite $H_{MBS}(t)$ and $H_{QD-M}(t)$ as:
\begin{equation}
H_{MBS}(t)=\varepsilon_{m}(t)(f^{\dag}f-\frac{1}{2}),
\end{equation}
\begin{equation}
\begin{split}
H_{QD-M}(t)=&\frac{1}{\sqrt{2}}[(V_1(t)+iV_2(t))d^{\dag}f^{\dag}\\
&+(V_1(t)-iV_2(t))d^{\dag}f\\
&-(V_1^{*}(t)+iV_2^{*}(t))df^{\dag}\\
&-(V_1^{*}(t)-iV_2^{*}(t))df].
\end{split}
\label{HQDM1}
\end{equation}
We introduce two representations: $\Phi^{\dag}=[d^{\dag}, d, f^{\dag}, f]$ and $D_{\eta k}^{\dag}=[c^{\dag}_{\eta k}, c_{\eta k}, 0, 0]$. Based on these, we can define four types of Green functions as follows:
\begin{equation}
G_{\Phi}^{\gamma}(t,t_1) = \langle \langle \Phi(t); \Phi^{\dag}(t_1) \rangle \rangle^{\gamma},
\end{equation}
\begin{equation}
G_{\Phi,\eta k}^{\gamma}(t,t_1) = \langle \langle \Phi(t); D_{\eta k}^{\dag}(t_1) \rangle \rangle^{\gamma},
\end{equation}
\begin{equation}
G_{\eta k, \Phi}^{\gamma}(t,t_1) = \langle \langle D_{\eta k}(t); \Phi^{\dag}(t_1) \rangle \rangle^{\gamma},
\end{equation}
\begin{equation}
G_{\eta k}^{\gamma}(t,t_1) = \langle \langle D_{\eta k}(t); D_{\eta k}^{\dag}(t_1) \rangle \rangle^{\gamma},
\end{equation}
where $\gamma$ can be ``$R, A, <, >$".

\subsection{Equations of motion}

To describe the dynamical behavior of the hybrid system, we introduce four density matrices: $\rho_d(t)=\left\langle d^{\dag}(t)d(t) \right\rangle$, representing the electron density of the QD; $\rho_f(t)=\left\langle f^{\dag}(t)f(t) \right\rangle$, representing the density of the MBSs; $\rho_1(t)=\left\langle f^{\dag}(t)d(t) \right\rangle$ and $\rho_2(t)=\left\langle f(t)d(t) \right\rangle$, which describe the tunneling process between the QD and the two MBSs. Their EOMs can be obtained from the Heisenberg equation of motion, incorporating the time-dependent NGF method:
\begin{equation}
\begin{split}
\dot{\rho}_{d}(t)=&\frac{1}{\sqrt{2}}\{[iV_1^{*}(t)-V_2^{*}(t)]\rho_{1}(t)+[iV_1^{*}(t)+V_2^{*}(t)]\rho_{2}(t)\\
&-[iV_1(t)+V_2(t)]\rho_{1}^{\dag}(t)-[iV_1(t)-V_2(t)]\rho_{2}^{\dag}(t)\}\\
&+\sum_{\eta}[B_{\eta,11}^{<}(t)+(B_{\eta,11}^{<}(t))^{\dag}],
\end{split}
\end{equation}
\begin{equation}
\begin{split}
\dot{\rho}_{f}(t)=&\frac{1}{\sqrt{2}}\{[-iV_1^{*}(t)+V_2^{*}(t)]\rho_{1}(t)+[iV_1^{*}(t)+V_2^{*}(t)]\rho_{2}(t)\\
&+[iV_1(t)+V_2(t)]\rho_{1}^{\dag}(t)-[iV_1(t)-V_2(t)]\rho_{2}^{\dag}(t)\},
\end{split}
\end{equation}
\begin{equation}
\begin{split}
\dot{\rho}_{1}(t)=&\frac{1}{\sqrt{2}}[iV_1(t)+V_2(t)](\rho_{d}(t)-\rho_{f}(t))\\
&+i(\varepsilon_m(t)-\varepsilon_d(t))\rho_1(t)-\sum_{\eta}B_{\eta,42}^{<}(t),
\end{split}
\end{equation}
\begin{equation}
\begin{split}
\dot{\rho}_{2}(t)=&\frac{1}{\sqrt{2}}[iV_1(t)-V_2(t)](\rho_{d}(t)+\rho_{f}(t)-1)\\
&-i(\varepsilon_m(t)+\varepsilon_d(t))\rho_2(t)-\sum_{\eta}B_{\eta,32}^{<}(t),
\end{split}
\end{equation}
where $B_{\eta}^{<}(t)=\sum_k V_{\eta k}(t)G_{\Phi,\eta k}^{<}(t,t)\hat{\alpha}_{1}$, and
\begin{equation}
\hat{\alpha}_{1}=\left[
\begin{matrix}
1 & 0 & 0 & 0\\
0 & -1 & 0 & 0\\
0 & 0 & 0 & 0\\
0 & 0 & 0 & 0
\end{matrix}
\right].
\end{equation}
By applying the Dyson equation and Langreth's theorem, we obtain:
\begin{equation}
\begin{split}
B_{\eta}^{<}(t)=&\int dt_1\sum_{k}V_{\eta k}(t)V_{\eta k}(t_1)[G_{\Phi}^{R}(t,t_1)\hat{\alpha}_{1}G_{\eta k}^{<}(t_1,t)\hat{\alpha}_{1}\\
&+G_{\Phi}^{<}(t,t_1)\hat{\alpha}_{1}G_{\eta k}^{A}(t_1,t)\hat{\alpha}_{1}]\\
=&\frac{i}{2}\Gamma_{\eta}(t)[G_{\Phi}^{R}(t,t)+G_{\Phi}^{<}(t,t)]\hat{\alpha}_{1}^{2}+\sum_{p}u_{\eta}(t)\hat{\Pi}_{\eta p},
\end{split}
\end{equation}
where $\hat{\Pi}_{\eta p}(t)$ is the auxiliary density matrices:
\begin{equation}
\begin{split}
\hat{\Pi}_{\eta p}(t)=&\int dt_1 u_{\eta}(t_1)\Gamma_{\eta}^{0}\frac{R_{p}}{\beta}G_{\Phi}^{R}(t,t_1)\\
&\times\left[
\begin{matrix}
e^{-i\int_{t}^{t_1}dt_2\chi_{\eta p}^{\dag}(t_2)} & 0 & 0 & 0\\
0 & e^{-i\int_{t_1}^{t}dt_2\chi_{\eta p}^{-}(t_2)} & 0 & 0\\
0 & 0 & 0 & 0\\
0 & 0 & 0 & 0
\end{matrix}
\right].
\end{split}
\label{adm}
\end{equation}
In this equation, we utilize the Pad$\acute{e}$ expansion for the Fermi distribution: $f_{\eta}(\varepsilon)=\frac{1}{2}-\sum_{p}\frac{R_{p}}{\beta}(\frac{1}{\varepsilon-\chi_{\eta p}^{\dag}}
+\frac{1}{\varepsilon-\chi_{\eta p}^{-}})$, where $\chi_{\eta p}^{\pm}=\mu_{\eta}\pm\frac{i\chi_{p}}{\beta}$ and $\beta=\frac{1}{T}$. Here, $T$ denotes the system temperature, $\mu_{\eta}$ is the chemical potential of the lead $\eta$, $R_{p}$ and $\chi_{p}$ can be given by the eigenvalue problem of the symmetric matrix $\hat{B}$,\cite{RN127}
\begin{equation}
\hat{B}|b_{p}\rangle = b_{p}|b_{p}\rangle,
\end{equation}
where $b_{p}$ is the eigenvalue of $\hat{B}$ and $|b_{p}\rangle$ is the eigenstate, with
\begin{equation}
\hat{B}_{n,n+1} = \frac{1}{2\sqrt{(2n-1)(2n+1)}},\,\,\,\,n \geq 1,
\end{equation}
as
\begin{equation}
\chi_{p} = \frac{1}{b_{p}},
\end{equation}
\begin{equation}
R_{p} = \frac{|\langle1|b_{p}\rangle|^{2}}{4b_{p}^{2}}.
\end{equation}
Utilizing Eq.~\eqref{adm}, we derive the EOM for $G_{\Phi}^{R}(t,t_1)$:
\begin{equation}
\begin{split}
i\frac{\partial}{\partial t}G_{\Phi}^{R}(t,t_1)=&\delta(t-t_1)+\Sigma_{M}^{R}(t)G_{\Phi}^{R}(t,t_1)\\
&+\sum_{\eta}A_{\eta}^{R}(t,t_1).
\end{split}
\end{equation}
Here $A_{\eta}^{R}(t,t_1)=\sum_{k}V_{\eta k}(t)\hat{\alpha_{1}}G_{\eta k,\Phi}^{R}(t,t_1)$, and
\begin{equation}
\Sigma_{M}^{R}(t)=\left[
\begin{matrix}
0 & A_v(t) \\
A_v^{\dag}(t) & 0
\end{matrix}
\right],
\end{equation}
with
\begin{equation}
A_v(t)=\frac{1}{\sqrt{2}}\left[
\begin{matrix}
[V_1(t)-iV_2(t)] & [V_1(t)+iV_2(t)]\\
[-V_1^{*}(t)+iV_2^{*}(t)] & -[V_1^{*}(t)+iV_2^{*}(t)]
\end{matrix}
\right].
\end{equation}
Upon reapplying the Dyson equation and Langreth's theorem, we obtain:
\begin{equation}
\begin{split}
A_{\eta}^{R}(t,t_1)=&\int dt_2\sum_{k}V_{\eta k}(t)V_{\eta k}(t_2)\hat{\alpha_{1}}G_{\eta k}^{R}(t,t_2)\hat{\alpha}_{1}G_{\Phi}^{R}(t_2,t_1)\\
=&-\frac{i}{2}\Gamma_{\eta}(t)\hat{\alpha}_{1}^{2}G_{\Phi}^{R}(t,t_1).
\end{split}
\end{equation}
Finally, we derive the EOMs for these density matrix elements and auxiliary density matrices:
\begin{equation}
\begin{split}
\dot{\rho}_{d}(t)=&\frac{1}{\sqrt{2}}\{[iV_1^{*}(t)-V_2^{*}(t)]\rho_{1}(t)+[iV_1^{*}(t)+V_2^{*}(t)]\rho_{2}(t)\\
&-[iV_1(t)+V_2(t)]\rho_{1}^{\dag}(t)-[iV_1(t)-V_2(t)]\rho_{2}^{\dag}(t)\}\\
&+\sum_{\eta p}u_{\eta}(t)[\Pi_{\eta p,11}(t)+\Pi_{\eta p,11}^{\dag}(t)]\\
&+\frac{1}{2}\Gamma(t)(1-2\rho_d(t)),
\end{split}
\label{EOM1}
\end{equation}
\begin{equation}
\begin{split}
\dot{\rho}_{f}(t)=&\frac{1}{\sqrt{2}}\{[-iV_1^{*}(t)+V_2^{*}(t)]\rho_{1}(t)+[iV_1^{*}(t)+V_2^{*}(t)]\rho_{2}(t)\\
&+[iV_1(t)+V_2(t)]\rho_{1}^{\dag}(t)-[iV_1(t)-V_2(t)]\rho_{2}^{\dag}(t)\},
\end{split}
\label{EOM2}
\end{equation}
\begin{equation}
\begin{split}
\dot{\rho}_{1}(t)=&\frac{1}{\sqrt{2}}[iV_1(t)+V_2(t)](\rho_{d}(t)-\rho_{f}(t))\\
&+(-\frac{1}{2}\Gamma(t)+i\varepsilon_m(t)-i\varepsilon_d(t))\rho_1(t)\\
&+\sum_{\eta p}u_{\eta}(t)\Pi_{\eta p,31}^{*}(t),
\end{split}
\label{EOM3}
\end{equation}
\begin{equation}
\begin{split}
\dot{\rho}_{2}(t)=&\frac{1}{\sqrt{2}}[iV_1(t)-V_2(t)](\rho_{d}(t)+\rho_{f}(t)-1)\\
&+(-\frac{1}{2}\Gamma(t)+i\varepsilon_m(t)+i\varepsilon_d(t))\rho_2(t)\\
&+\sum_{\eta p}u_{\eta}(t)\Pi_{\eta p,41}^{*}(t),
\end{split}
\label{EOM4}
\end{equation}
\begin{equation}
\begin{split}
i\frac{\partial}{\partial t}\hat{\Pi}_{\eta p}(t)=&[\Sigma_{M}^{R}(t)-\frac{i}{2}\Gamma(t)\hat{\alpha}_1^2-\varepsilon_d(t)\hat{\alpha}_1+\hat{\chi}_{\eta p}(t)]\hat{\Pi}_{\eta p}(t)\\
&+u_{\eta}(t)\Gamma_{\eta}^0\frac{R_p}{\beta}\hat{\alpha}_1^2,
\end{split}
\label{EOM5}
\end{equation}
where $\Gamma(t) = \Gamma_L(t)+\Gamma_R(t)$, and
\begin{equation}
\hat{\chi}_{\eta p}(t)=\left[
\begin{matrix}
-\chi_{\eta p}^{\dag}(t) & 0 & 0 & 0\\
0 & \chi_{\eta p}^{-}(t) & 0 & 0\\
0 & 0 & 0 & 0\\
0 & 0 & 0 & 0
\end{matrix}
\right].
\end{equation}

\subsection{Conductance, capacitance and relaxation resistance}

The electric current flowing from the left lead into the QD can be determined from the rate of change of the electron number operator for the left lead:
\begin{equation}
I_{L}(t)=-e\frac{d}{dt}\sum_{k}\langle c_{L k}^{\dag}(t)c_{L k}(t)\rangle.
\end{equation}
By expressing $I_{L}(t)$ in terms of the density matrix elements and auxiliary density matrices, the current flowing from the left lead can be rewritten as a more convenient form for calculation:
\begin{equation}
I_{L}(t)=\frac{1}{2}\Gamma_{L}(t)-\Gamma_{L}(t)\rho_{d}(t)+u_{L}(t)\sum_{p}[\Pi_{L p,11}(t)+\Pi^{*}_{L p,11}(t)].
\end{equation}
Therefore, by seeting $\mu_L$=$\frac{V}{2}$ and $\mu_R$=$-\frac{V}{2}$, and assuming that all system parameters are time-independent, we can evaluate the linear conductance of the system as the derivative of the current of the left lead, with respect to the voltage $V$ at zero voltage,
\begin{equation}
G=\frac{\partial I_L}{\partial V}\bigg|_{V=0}.
\label{G}
\end{equation}

To examine the linear adiabatic response of the hybrid system, we proceed under the assumption, without loss of generality, that a small low-frequency signal is applied to the QD. This results in the dot's energy level being modeled as $\varepsilon_d(t)$ = $\varepsilon_d + \varepsilon_{ac}(t)$, where the time-dependent component is given by $\varepsilon_{ac}(t)=\varepsilon_{ac0}\sin(\theta)$ with $\theta = \omega t$ and $\omega$ denoting the tuning frequency. To satisfy the adiabatic condition, $\omega$ must be set to zero. Furthermore, we assume that all other system parameters remain time-independent.
In the literature, two key physical quantities are introduced to characterize the adiabatic response current, $I(t) = \partial_{t}\rho_d(t)$, of the mesoscopic circuit: the differential capacitance, $C_{\partial}(t)$, and the differential relaxation resistance, $R_{\partial}(t)$. The current $I(t)$ can thus be expressed as\cite{RC1}:
\begin{equation}
I(t) = - C_{\partial}(t) \partial_{t} \varepsilon_{ac}(t) + R_{\partial}(t) C_{\partial}(t) \partial_{t} (C_{\partial}(t)\partial_{t}\varepsilon_{ac}(t)).
\label{RCtime}
\end{equation}
It is important to recognize that these two quantities are primarily believed to be governed by the first-order and second-order processes of single-electron tunneling through the QD, respectively.
Furthermore, in this study, we focus on the linear transport regime, which involves the transient response to an extremely small amplitude of time modulation, particularly as $\varepsilon_{ac0}\to 0$. Therefore, in the linear transport regime, we can expand the transient current $I(t)$, Eq.~\eqref{RCtime}, up to the second order with respect to the driving voltage $\varepsilon_{ac}(t)$ as
\begin{equation}
\omega \dot{\rho}_{d}(\theta) = -\omega C_{d}\dot{\varepsilon}_{ac}(\theta)+\omega^{2}C^{2}_{d}R_{d}\ddot{\varepsilon}_{ac}(\theta),
\label{RC}
\end{equation}
where `` $\dot{}$ '' denotes the derivative with respect to $\theta$.
Interestingly, the two response functions, $C_{\partial}(t)$ and $R_{\partial}(t)$, become time-independent, manifesting as the adiabatic linear capacitance $C_{\partial}(t) = C_d$, and the adiabatic linear relaxation resistance $R_{\partial}(t) = R_d$.

To calculate the two adiabatic quantities, we recast the resulting EOMs (Eqs.~\eqref{EOM1}-\eqref{EOM5}) into a compact matrix form:
\begin{equation}
\omega \dot{\psi}(\theta) = \hat{A}(\theta)\psi(\theta)+\hat{B},
\label{EOM_change}
\end{equation}
where $\psi(\theta) = [\rho_{d}(\theta), \rho_{f}(\theta), \rho_{1}(\theta), \rho_{2}(\theta), \hat{\Pi}_{\eta p,1}(\theta), \hat{\Pi}_{\eta p,1}^{*}(\theta)$, $\hat{\Pi}_{\eta p,2}(\theta), \hat{\Pi}_{\eta p,2}^{*}(\theta)]^{T}$, and $\hat{\Pi}_{\eta p,i}(\theta)$ = $[\hat{\Pi}_{\eta p,1i}(\theta)$, $\hat{\Pi}_{\eta p,2i}(\theta)$, $\hat{\Pi}_{\eta p,3i}(\theta)$, $\hat{\Pi}_{\eta p,4i}(\theta)]^{T}$ $(i = 1,2)$. The term $\hat{A}(\theta)$ denotes the dynamical matrix, while $\hat{B}$ represents the constant driving terms. Both $\hat{A}(\theta)$ and $\hat{B}$ can be directly obtained from the equations \eqref{EOM1}-\eqref{EOM5}.
Under the adiabatic condition, the time-varying auxiliary density matrices $\psi(\theta)$ can be expanded as a power series in the frequency $\omega$ as:
\begin{equation}
\psi(\theta) = \psi^{(0)}(\theta) + \omega \psi^{(1)}(\theta) + O(\omega^2).
\end{equation}
Therefore, by substituting this expansion into the time evolution equation \eqref{EOM_change}, we derive the hierarchical equations that dictate the behavior of those terms at different orders in $\omega$:
\begin{equation}
0 = \hat{A}(\theta)\psi^{(0)}(\theta) + \hat{B},
\end{equation}
and
\begin{equation}
\dot{\psi}^{(0)}(\theta) = \hat{A}(\theta)\psi^{(1)}(\theta).
\end{equation}
Then we have $\psi^{(0)}(\theta) = -\hat{A}^{-1}(\theta)B$, $\psi^{(1)}(\theta) = \hat{A}^{-2} (\theta) \dot{\hat{A}}(\theta) \hat{A}^{-1}(\theta)B$, and their derivatives, $\dot{\psi}^{(0)}(\theta)$ and $\dot{\psi}^{(1)}(\theta)$. At the end, from Eq.~\eqref{RC}, we obtain the adiabatic linear capacitance, $C_d$, and relaxation resistance, $R_d$, of the system, respectively, as:
\begin{equation}
C_d = -\frac{\dot{\rho}^{(0)}_{d}(\theta)}{\dot{\varepsilon}_{ac}(\theta)},
\label{Cd}
\end{equation}
and
\begin{equation}
R_d = (C_d)^{-2}\frac{\dot{\rho}^{(1)}_{d}(\theta)}{\ddot{\varepsilon}_{ac}(\theta)},
\label{Rd}
\end{equation}
with $\dot{\rho}^{(0)}_{d}(\theta)=[\dot{\psi}^{(0)}(\theta)]_1$ and $\dot{\rho}^{(1)}_{d}(\theta)=[ \dot{\psi}^{(1)}(\theta)]_1$.

\section{Results and Discussions}

In this section, we will employ the resulting formulations, Eqs.~\eqref{G}, ~\eqref{Cd}, and ~\eqref{Rd}, to systematically investigate the stationary transport and adiabatic response properties of the QD-MBSs hybrid system. For our calculations, we focus on a symmetric system where $\Gamma_{L}^0$ = $\Gamma_{R}^0$ = $\Gamma$, with $\Gamma$ adopted as the energy unit throughout this work. Without loss of generality, the tunneling parameters between the MBSs and the QD are defined as $V_1(t)$ = $1$ and $V_2(t)$ = $e^{i\phi}$, where $\phi$ represents the phase difference between the two MBSs.
The phase convention employed in our analysis maintains consistency with the symmetric phase selection, explicitly expressed as $V_1 = e^{-i\phi/2}$ and $V_2 = e^{i\phi/2}$, which is justified by performing the operators transformation $d= e^{i\phi/2}\tilde{d}$, $c_{\eta k}= e^{i\phi/2}\tilde{c}_{\eta k}$. This transformation yields the following modified Hamiltonian:
\begin{equation}
H_{QD}(t)=\varepsilon_d(t)\tilde{d}^{\dag}\tilde{d},
\end{equation}
\begin{equation}
H_{QD-M}=e^{-i\frac{\phi}{2}}\tilde{d}^{\dag}\gamma_1+e^{i\frac{\phi}{2}}\gamma_1\tilde{d}+e^{i\frac{\phi}{2}}\tilde{d}^{\dag}\gamma_2+e^{-i\frac{\phi}{2}}\gamma_2\tilde{d},
\end{equation}
\begin{equation}
H_{Lead}=\sum_{\eta k}\varepsilon_{\eta k}\tilde{c}_{\eta k}^{\dag}\tilde{c}_{\eta k},
\end{equation}
\begin{equation}
H_{T}=\sum_{\eta k}V_{\eta k}(\tilde{c}_{\eta k}^{\dag}\tilde{d}+\tilde{d}^{\dag}\tilde{c}_{\eta k}).
\end{equation}
This result aligns precisely with the Hamiltonian derived using the symmetric phase convention where $V_1 = e^{-i\phi/2}$ and $V_2 = e^{i\phi/2}$.

The system temperature is fixed at $0.01\Gamma$, unless otherwise specified.

\subsection{Linear conductance}

First, we examine the case where the phase difference between the two MBSs is zero. Figure 2(a) shows the linear conductance as a function of the bare dot level $\varepsilon_d$, comparing the behavior for systems with different MBSs modes and without MBSs.
In the absence of MBSs, the linear conductance attains its upper bound, denoted as $G_0$, where $G_0$ is equal to $e^2/h$. However, when the MBSs are in the zero mode, the conductance is reduced to half of $G_0$, specifically to $G_0/2$, which is a distinctive characteristic of systems harboring MBSs. With an increase in the parameter $\varepsilon_m$, the linear conductance also exhibits an upward trend. This phenomenon occurs because, under such circumstances, the MBSs are composed not only of electron-hole pairs but also include single electrons or holes.

This characteristic is further illustrated in Fig. 3, which depicts the density of the MBSs, $\rho_f=\langle f^{\dag}f \rangle$, as a function of the bare dot level $\varepsilon_d$ for different MBSs modes. When the MBSs are in the zero mode, $\rho_f$ is 0.5,  indicating that each electron in the MBSs pairs with a hole to form an electron-hole pair. As $\varepsilon_m$ increases, $\rho_f$ decreases, signifying the presence of single holes in the MBSs, which in turn increases the linear conductance.

Figure 2(b) shows the linear conductance as a function of the bare dot level $\varepsilon_d$ under a phase difference of $\pi/2$ between the two MBSs, comparing the behavior for systems with different MBSs modes and without MBSs. Here, we observe that when the MBSs are in the zero mode, the linear conductance vanishes. In contrast, the conductance reaches $G_0$ when $\varepsilon_d$ satisfies the condition $\varepsilon_d = \frac{2}{\varepsilon_m}$.
\begin{figure}[h]
\centering
\includegraphics[scale=0.21]{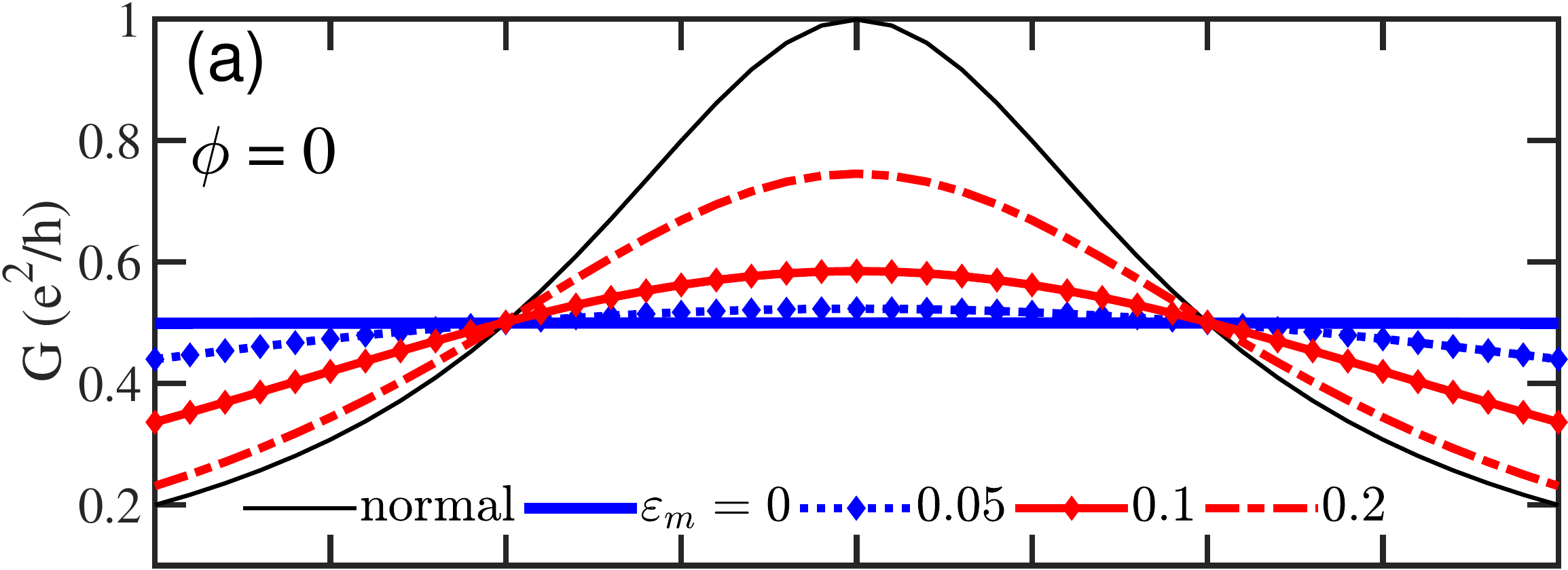}
\includegraphics[scale=0.21]{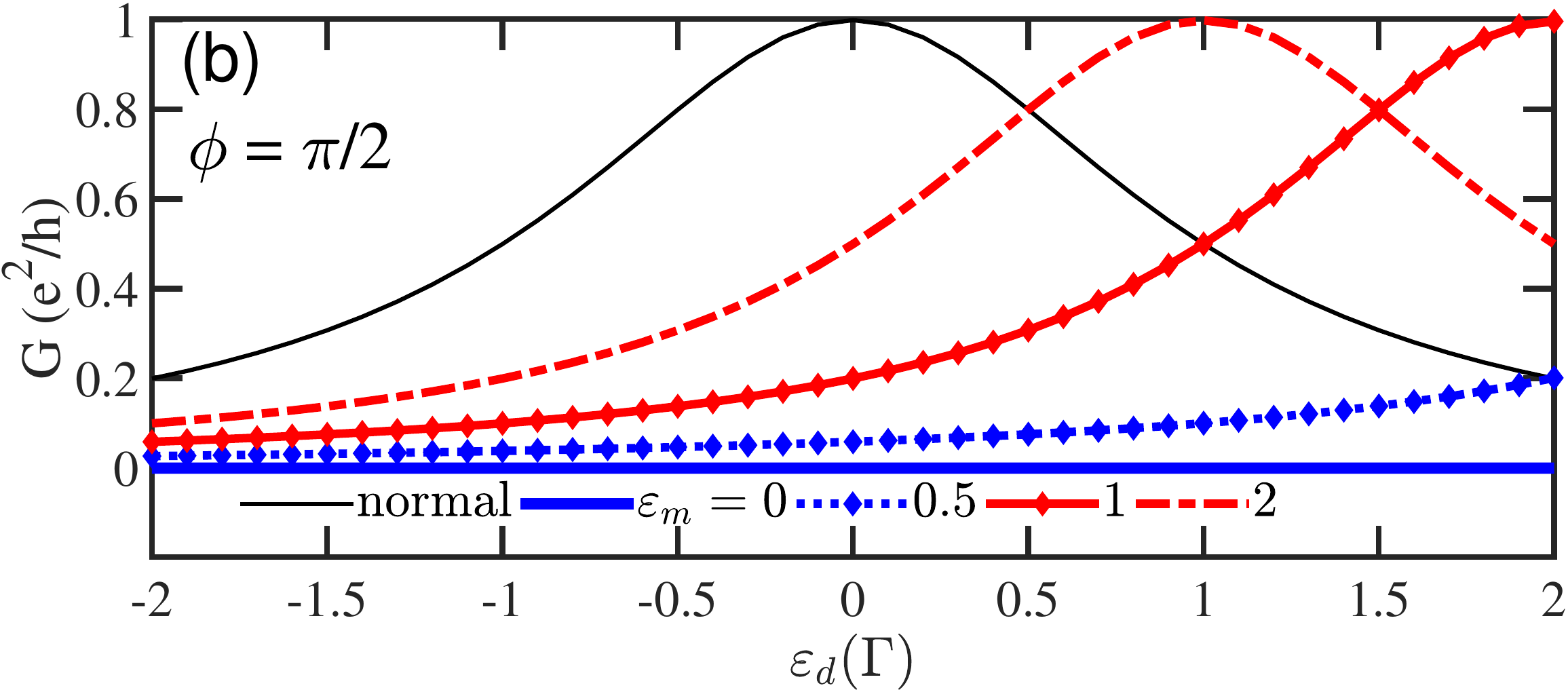}
\caption{(Colour online) The calculated linear conductance is plotted as a function of the bare dot level $\varepsilon_{d}$ for various modes of the MBSs with two different phase differences between the two MBSs, $\phi=0$ (a) and $\pi/2$ (b), respectively. For the sake of comparison, the linear conductance of the system in the absence of MBSs is also depicted.}.
\end{figure}

\begin{figure}[h]
\centering
\includegraphics[scale=0.21]{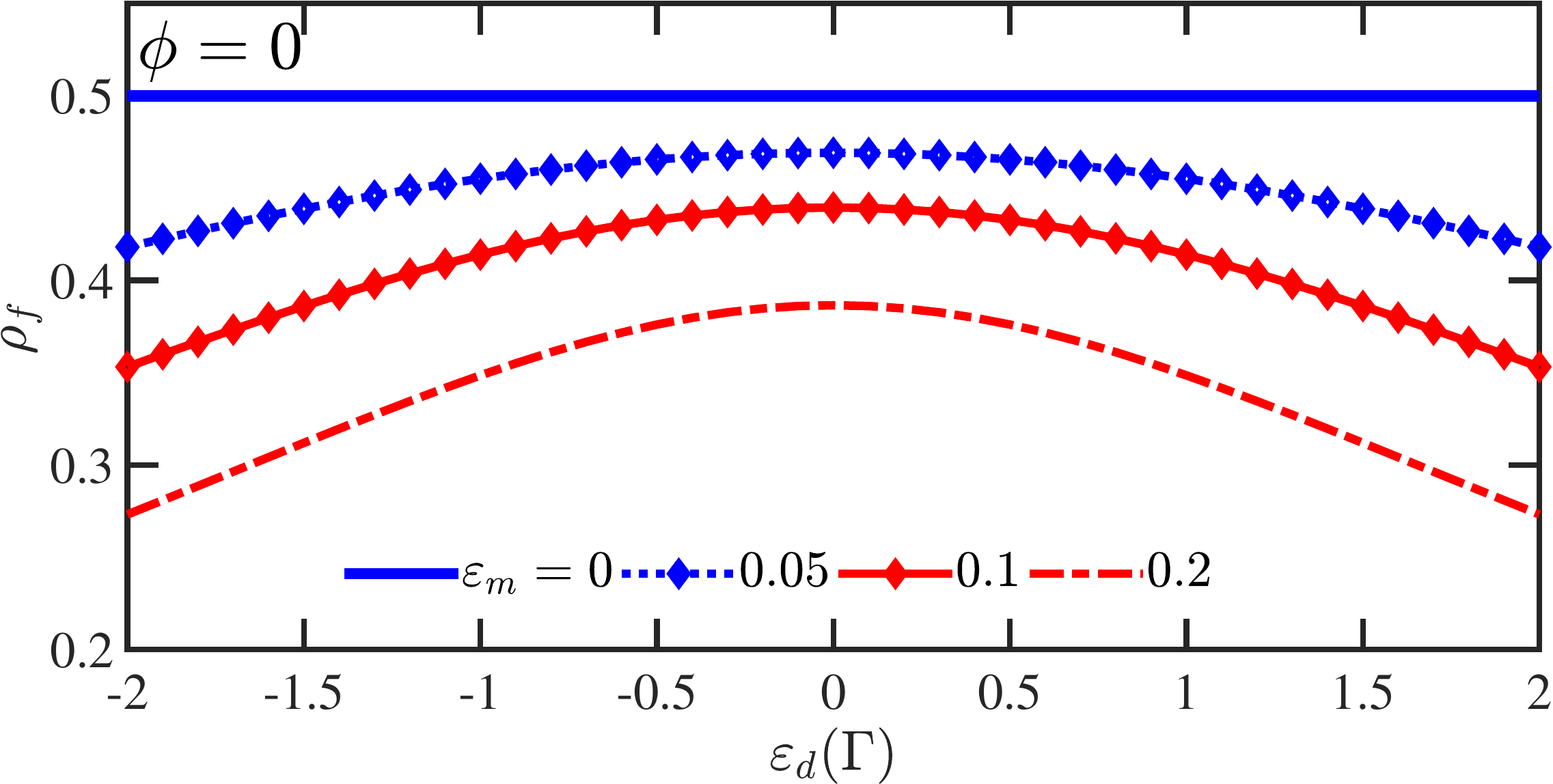}
\caption{(Colour online) The occupation number density of MBSs as a function of the dot level $\varepsilon_d$ for various modes of the MBSs, $\varepsilon_m$, with $\phi=0$.}.
\end{figure}

In the subsequent analysis, we investigate the linear conductance $G$ as it varies with the phase difference $\phi$ between the two MBSs, as depicted in Fig. 4, for different MBSs modes. Our findings reveal that the bare dot level of the QD significantly affects the periodicity of the conductance. As illustrated in Fig. 4(a), under the case of particle-hole symmetry, i.e. $\varepsilon_d=0$, the phase-dependent conductance displays a periodicity of $\pi$. In contrast, when the QD is shifted away from the particle-hole symmetric point, i.e. $\varepsilon_d \neq 0$, and $\varepsilon_m \neq 0$ the conductance exhibits a periodicity of $2\pi$ with respect to the phase difference, as shown in Fig. 4(b). Additionally, it is observed that when the MBSs are in the zero mode and the phase difference is not equal to an integer multiple of $\pi$, the linear conductance is completely suppressed in both scenarios.
This phenomenon bears resemblance to the flux-dependent conductance modulation observed in Aharonov-Bohm rings, where both systems demonstrate either partial conductance suppression or total blockade. The crucial distinction lies in their respective control mechanisms: while Aharonov-Bohm rings achieve complete conductance suppression only at discrete flux values, our system exhibits total conductance blockade for all phase differences except $0$ and $\pi$ when satisfies $\varepsilon_{m} = 0$, demonstrating a broader operational range for conductance control.
Moreover, Fig. 4(b) demonstrates that the system's linear conductance undergoes a complete inversion—transitioning from local minima to global maxima—under the parameter conditions of $\epsilon_d = \Gamma$ with $\epsilon_m \in \{2, 5\}$ and $\phi \in \{-1.5\pi, 0.5\pi\}$.
\begin{figure}[h]
\centering
\includegraphics[scale=0.21]{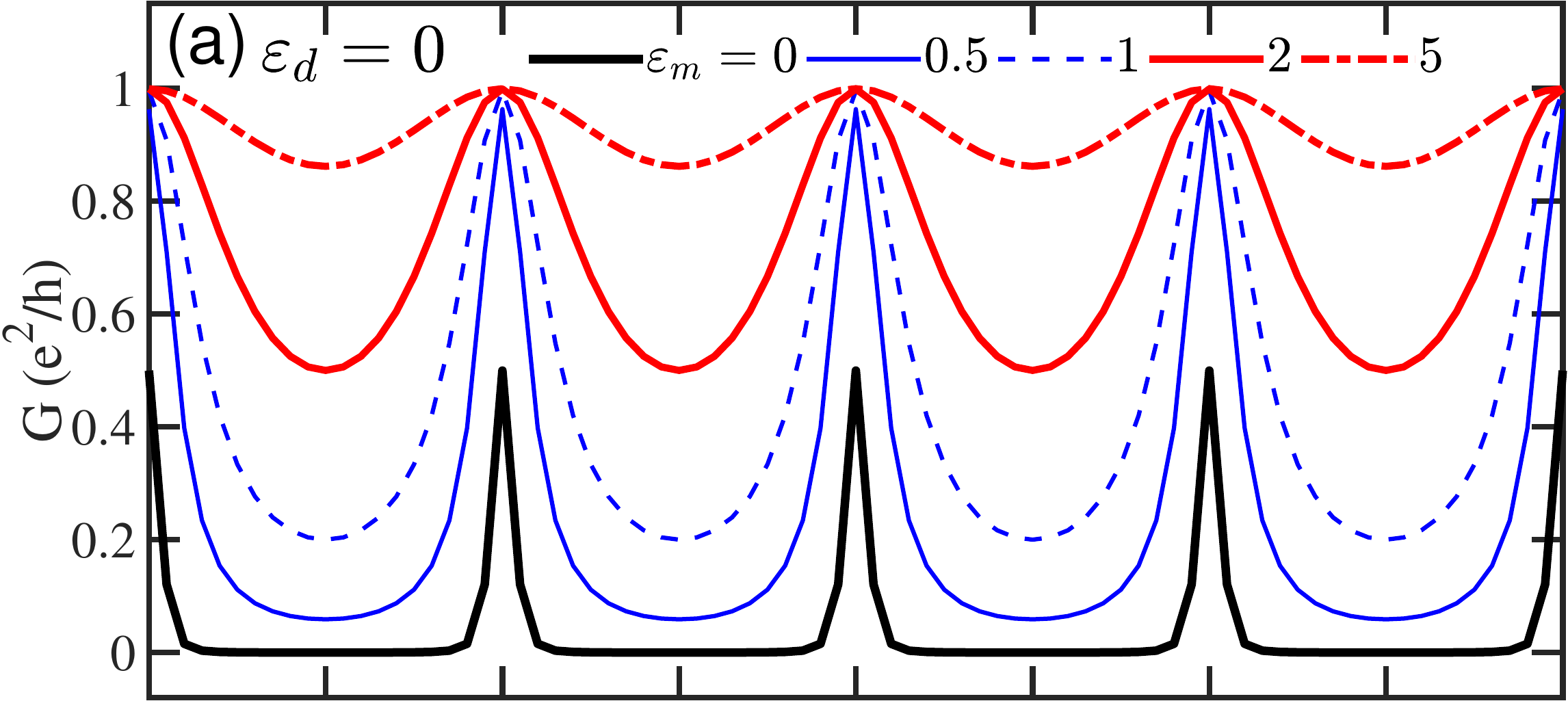}
\includegraphics[scale=0.21]{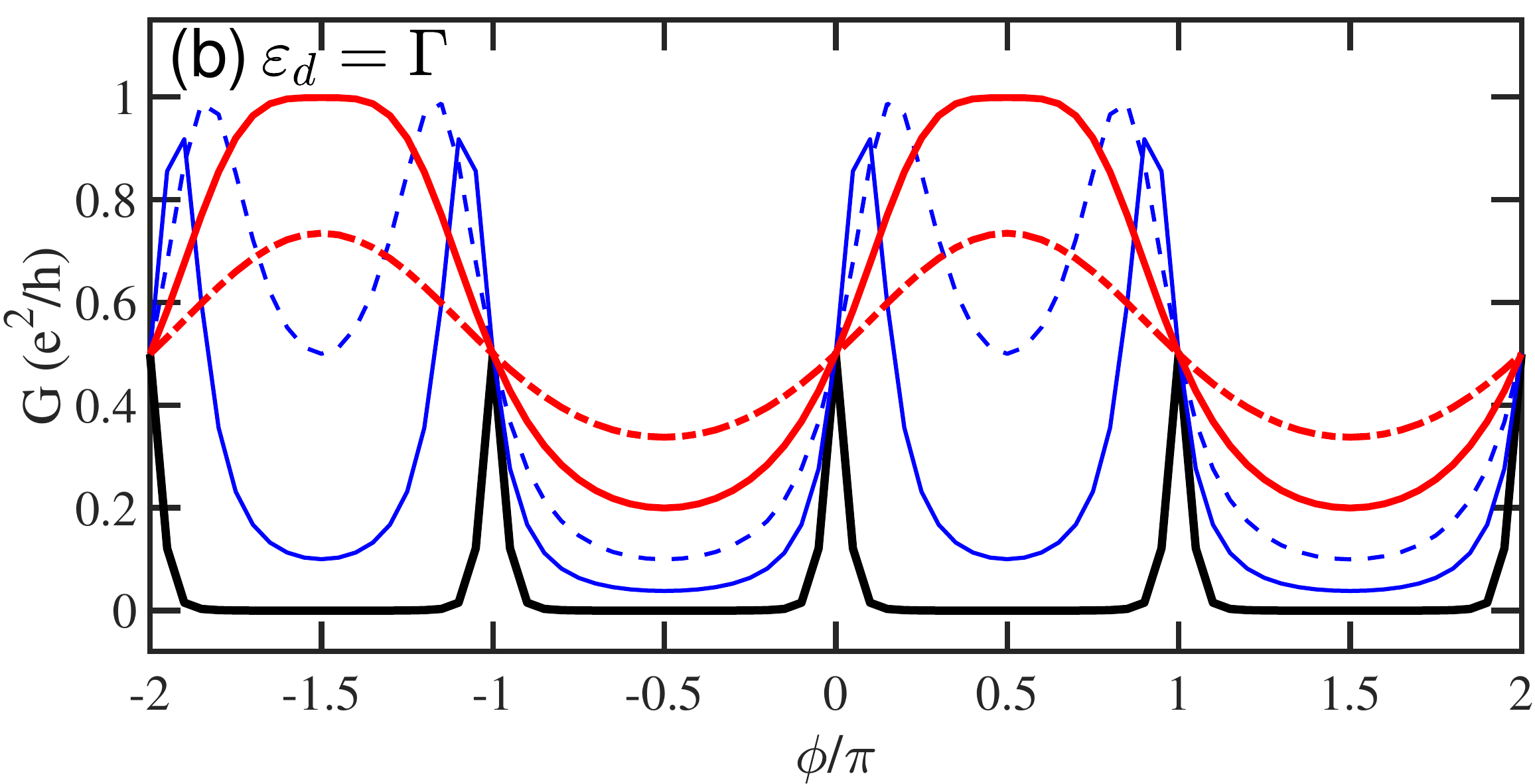}
\caption{(Colour online) The linear conductance as a function of the phase difference $\phi$ between the two MBSs for various modes of MBSs, with two specific dot levels $\varepsilon_d=0$ (a) and $\Gamma$ (b), respectively.}.
\end{figure}

To interpret these calculated results as shown in Fig. 4, we further rewrite the Hamiltonian of the MBSs. In the case of $\phi \neq n\pi$, it is convenient to introduce a new fermion operator $\tilde{f}$ to depict the MBSs,
\begin{equation}
\gamma_1 = \frac{\tilde{f}+\tilde{f}^{\dag}}{\sqrt{2}},
\end{equation}
\begin{equation}
\gamma_2 = -\frac{e^{i\phi}\tilde{f}+e^{-i\phi}\tilde{f}^{\dag}}{\sqrt{2}}.
\end{equation}
Without loss of generality, we assume $V_2 = e^{i\phi}V_1$. By using the new operator, the Hamiltonians of the MBSs and tunneling process between the QD and the MBSs, $H_{MBS}$ and $H_{QD-M}$, then can be transformed as:
\begin{equation}
\begin{split}
H_{MBS}&=\varepsilon_m \sin\phi \tilde{f}^{\dag}\tilde{f}-\frac{i}{2}\varepsilon_{m}e^{-i\phi},\\
H_{QD-M}&=\frac{V_1}{\sqrt{2}}[(1-e^{-2i\phi})d^{\dag}\tilde{f}+(1-e^{2i\phi})\tilde{f}^{\dag}d].
\end{split}
\end{equation}
Different from Eq.~\eqref{HQDM1}, we observe that the hopping Hamiltonian no longer incorporates the anomalous pairing terms, $f^{\dag}d^{\dag}$ and $fd$. It is noticed that this transformed Hamiltonian resembles that of a Fano system, consisting of a non-interacting QD connected to two normal leads and another non-interacting QD. Therefore, by employing the NGF method within the context of the transformed model, we can readily deduce the linear conductance as follows\cite{RN27}:
\begin{equation}
G=\frac{\varepsilon_m^2\Gamma_L\Gamma_R}{[(\varepsilon_d\varepsilon_m-2V^2_1 \sin\phi)^2+\frac{1}{4}\varepsilon_m^2(\Gamma_L+\Gamma_R)^2]}.
\end{equation}
It is evident that the linear conductance $G$ completely vanishes if the MBSs are in the zero mode, i.e. $\varepsilon_m=0$.
Moreover, when $\varepsilon_d = 0$, the conductance demonstrates a $\pi$-periodic dependence on the phase difference $\phi$ between the MBSs, and increases monotonically with $\varepsilon_m$. In contrast, for the cases of $\varepsilon_d\neq 0$, the periodicity shifts to $2\pi$, and the conductance exhibits a non-monotonic behavior. Notably, when $\varepsilon_d$ satisfies the condition $\varepsilon_d = \frac{2V^2_1 \sin\phi}{\varepsilon_m}$, the system manifests a zero-bias conductance peak, which can be expressed as $G_P = \frac{4\Gamma_L\Gamma_R}{(\Gamma_L+\Gamma_R)^2}$.
Under the condition where $\epsilon_d = \Gamma_L = \Gamma_R = \Gamma$, $V_1 = 1$ and $\epsilon_m>0$, the conductance of the system can be analytically obtained from Eq. (52) and expressed in the following form:
\begin{equation}
G=\frac{1}{(1-\frac{2\sin\phi}{\epsilon_m})^2+1},
\end{equation}
evidently, for $\epsilon_m<2$ (corresponding to $2/\epsilon_m>1$), the conductance attains its peak value when the phase difference $\phi$ satisfies $\sin\phi = \epsilon_m/2$, while decreasing to local minima as $\phi$ approaches $-1.5\pi$ or $0.5\pi$ due to the growing $\sin\phi$ term. In contrast, when $\epsilon_m\geq2$ (where $2/\epsilon_m\leq1$), the conductance exhibits a monotonic increase with $\sin\phi$, reaching its maximum values precisely at $\phi\in\{-1.5\pi,0.5\pi\}$.

\subsection{Linear capacitance and linear relaxation resistance}

In the following, we discuss the linear capacitance $C_d$ (as expressed in Eq.~\eqref{Cd}) and the linear relaxation resistance $R_d$ (as given in Eq.~\eqref{Rd}).
Figures 5 and 6 present our calculated results, $C_d$ and $R_d$, respectively, plotted as functions of the dot level $\epsilon_d$ across various MBS modes.
For comparison, we also plot the corresponding results obtained in the absence of MBSs.

It is already known that the linear capacitance $C_d$ for a conventional system attains its maximum value, nearly $2e^2/\Gamma$, when $\epsilon_d=0$.\cite{RC1}
Our calculations reveal that in the presence of the zero mode of MBSs, this maximum capacitance value is reduced by half, to roughly $0.8e^2/\Gamma$ at $\epsilon_d=0$.
This reduction in capacitance can be attributed to the competition between single-electron tunneling and a novel tunneling process that emerges when a QD is coupled to MBSs.
When the QD forms a hybrid system with zero mode MBSs, two distinct first-order tunneling processes emerge: a single-electron transfer process and an additional first-order process involving dynamically generated electron-hole pairs. Crucially, while the electron-hole pair transfer between the QD and leads preserves the electron number in the QD, this process makes no contribution to the capacitance. Therefore, the capacitance arises exclusively from the single-electron tunneling channel. In this configuration, the existence of electron-hole pair tunneling effectively suppresses the probability amplitude of the single-electron process (since the combined probabilities of both processes must satisfy the normalization condition), consequently resulting in a measurable reduction of the system's capacitance.
The relationship between the energy $\varepsilon_m$ of the MBSs and the linear capacitance is inversely proportional:
as $\epsilon_m$ increases, the number of electron-hole pairs within the QD diminishes, which correspondingly suppresses the occurrence probability of first-order electron-hole pair tunneling processes while simultaneously amplifying the single-particle tunneling probability. Given that capacitance arises exclusively from single-particle tunneling contributions, this mechanism directly results in a enhancement of the system's capacitance.
Furthermore, the phase difference $\phi$ influences the position of the peak in the capacitance, as illustrated in Fig. 5(b). Specifically, the peak of the linear capacitance shifts to a position where $\epsilon_d$ is approximately inversely proportional to $\varepsilon_m$.
\begin{figure}[h]
\centering
\includegraphics[scale=0.21]{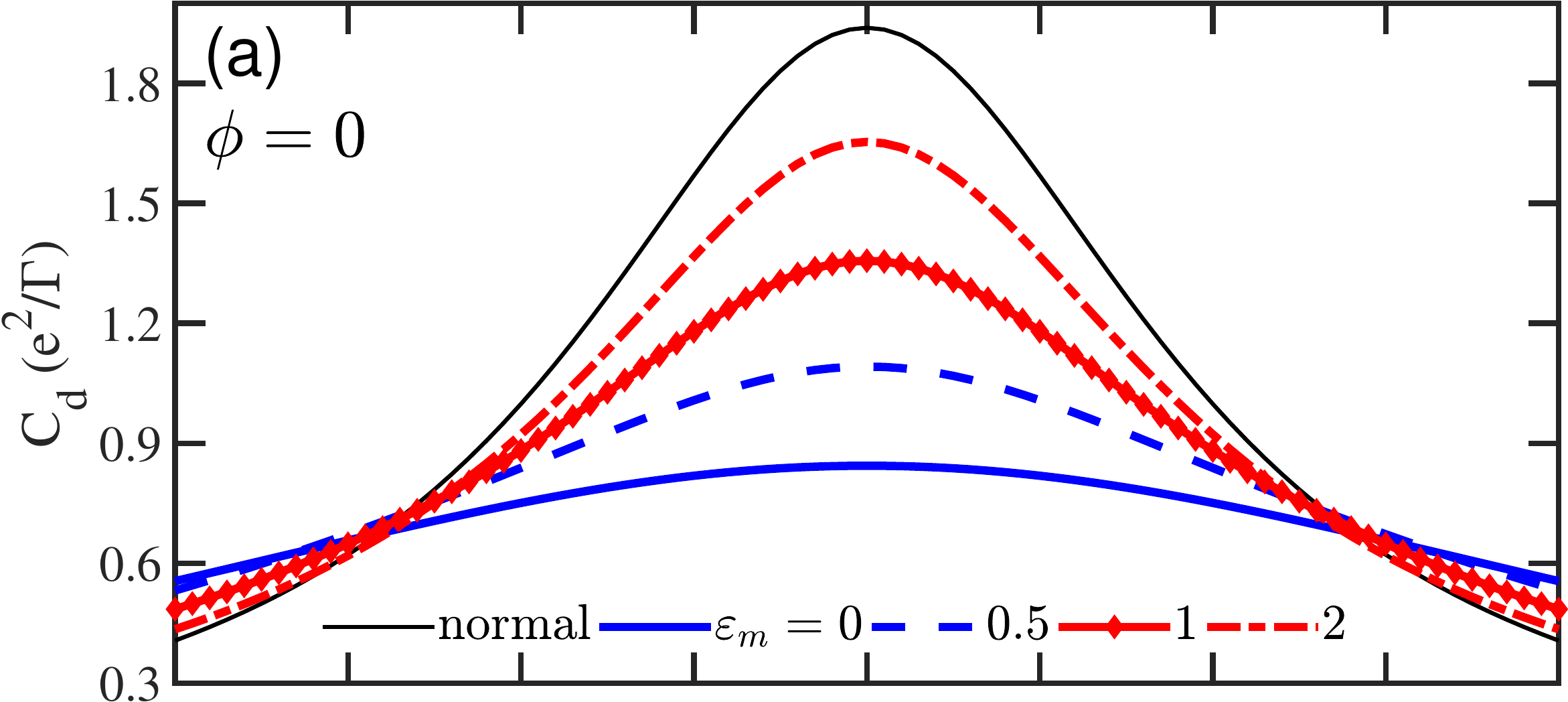}
\includegraphics[scale=0.21]{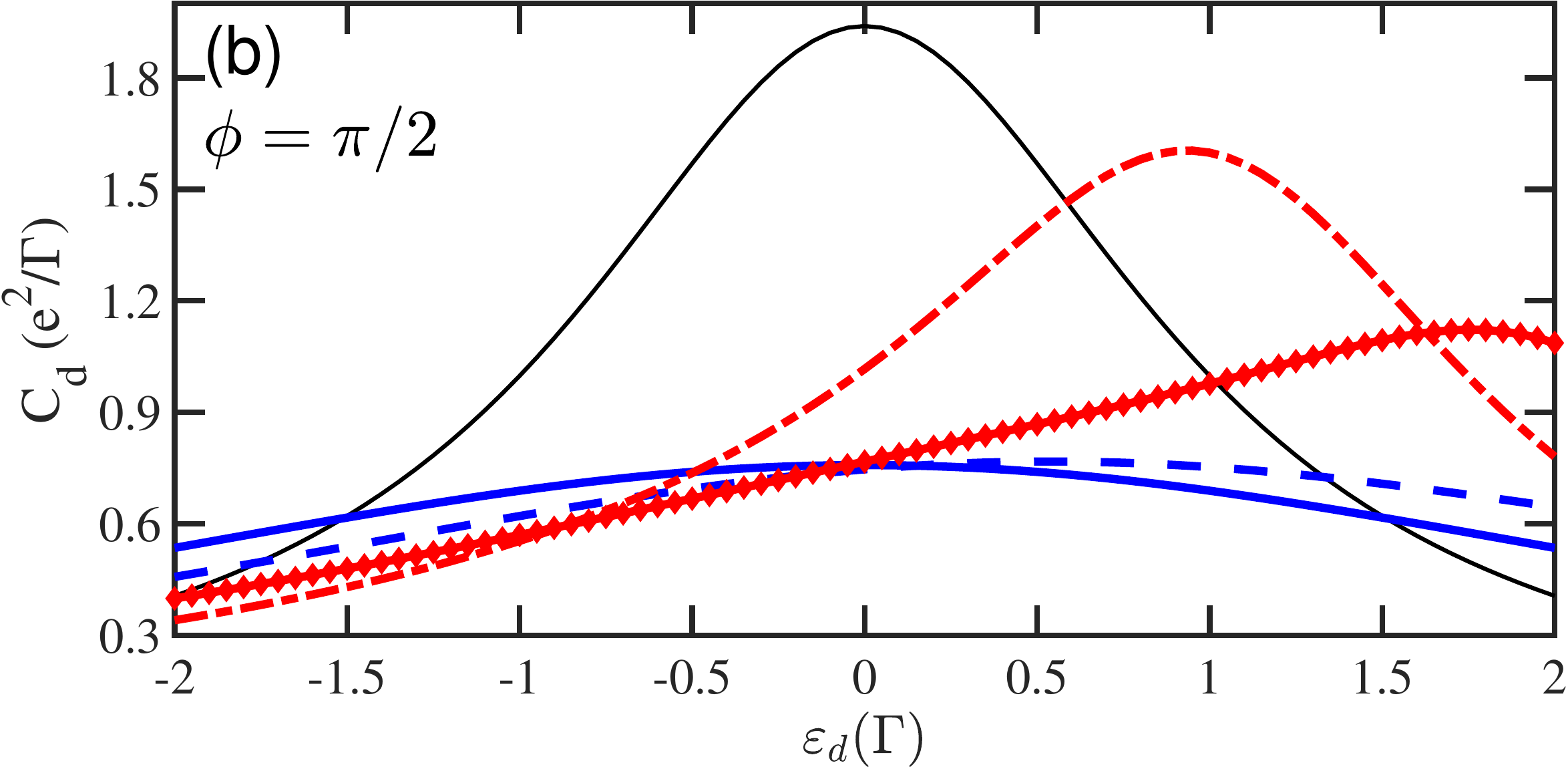}
\caption{(Colour online) The calculated linear capacitance $C_d$ is plotted as a function of the dot level $\varepsilon_d$ for various modes of MBSs with two distinct phase differences, $\phi=0$ (a) and $\pi/2$ (b), respectively. For the sake of comparison, the corresponding result for the system without MBSs is also shown.}
\end{figure}

Unlike the capacitance, the linear resistance is mainly governed by second-order tunneling processes occurring through the QD. In the context of the QD-Majorana hybrid system, two distinct types of second-order tunneling processes are identified: (1) charge-conserving processes described by the terms $d^{\dag}f$ and $df^{\dag}$ in the Hamiltonian $H_{QD-M}$ [Eq.~\eqref{HQDM1}] and (2) Cooper-pairing-like processes characterized by the terms $d^{\dag}f^{\dag}$ and $df$.
At the phase point $\phi=0$ where $V_1=V_2=1$, the quantum tunneling amplitudes take specific values: $1-i$ for the $d^{\dag}f$ process, $-(1+i)$ for $df^{\dag}$, $1+i$ for $d^{\dag}f^{\dag}$, and $-(1-i)$ for $df$. This configuration leads to an exact cancellation between the $d^{\dag}f$ and $df$ processes due to their identical magnitudes but opposite signs, while the $df^{\dag}$ and $d^{\dag}f^{\dag}$ processes similarly offset each other through their equal-and-opposite quantum amplitudes.
As a result, the charge-conserving processes and Cooper-pairing-like processes precisely counterbalance each other, leading to a vanishing linear resistance in the case of zero mode of MBSs, as illustrated in Fig.~6(a).\cite{RN36,RN8} It is worth noting that in the absence of MBSs, the linear resistance is approximately 0.5$h/e^2$. Moreover, when $\varepsilon_m \neq 0$, the non-zero coupling between the two ending MBSs can prevent these processes from completely canceling each other, thereby resulting in a non-zero linear resistance.\cite{RN8}

\begin{figure}[h]
\centering
\includegraphics[scale=0.21]{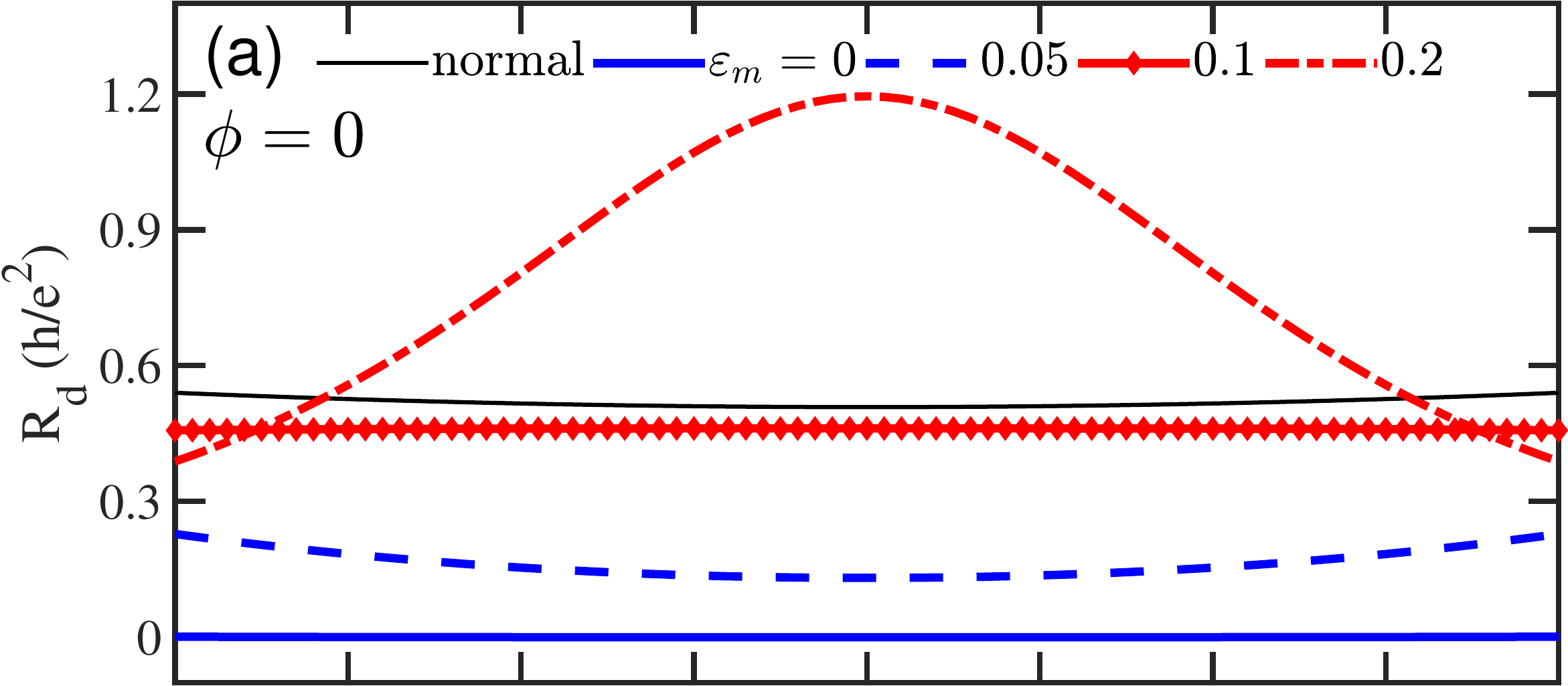}
\includegraphics[scale=0.21]{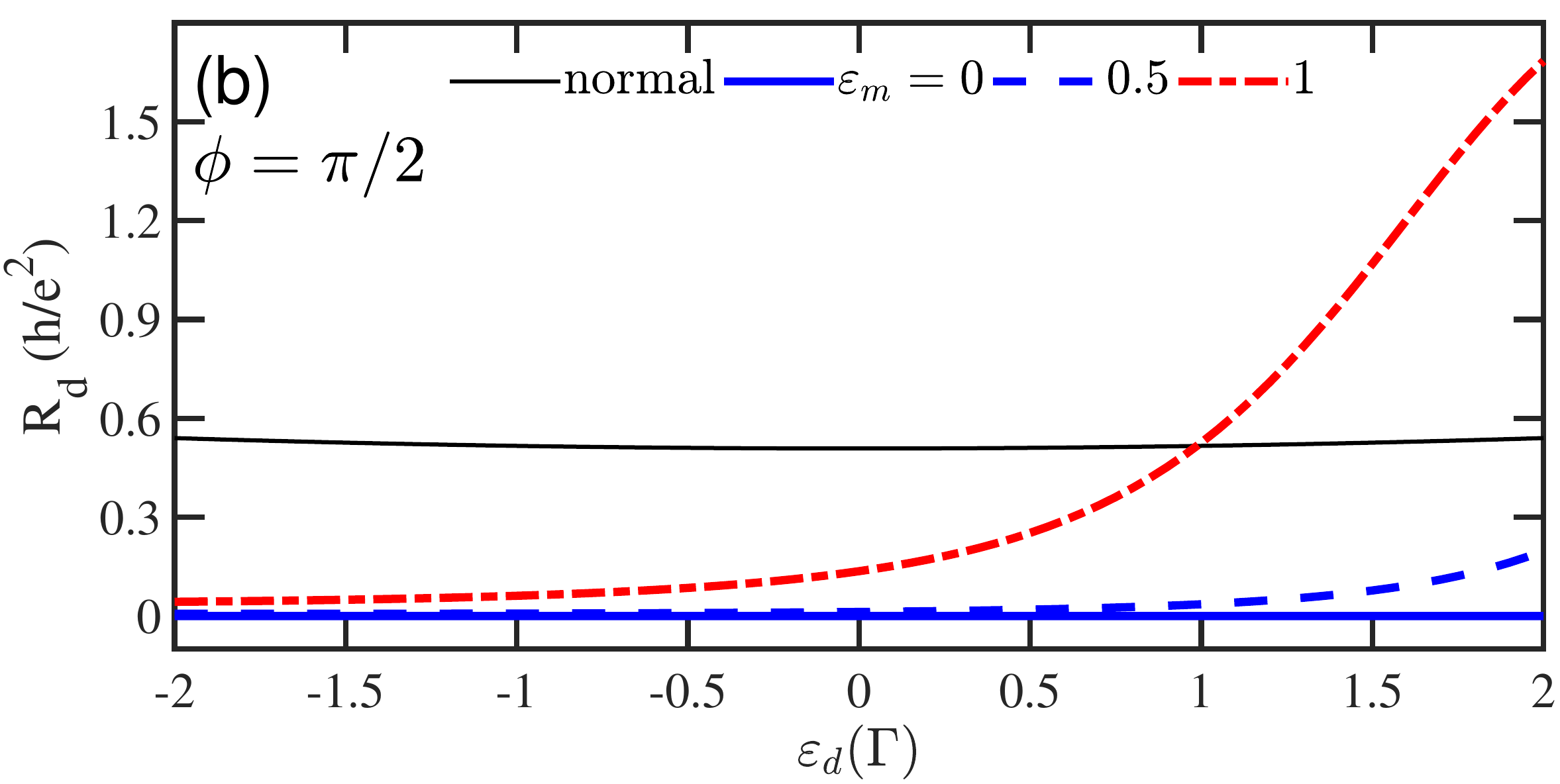}
\caption{(Colour online) The linear relaxation resistance $R_d$ is shown as a function of dot level with various modes of MBSs. The other parameters are the same as those in Fig. 5.}
\end{figure}

Nevertheless, the linear relaxation resistance exhibits a distinct behavior when the phase difference between the two MBSs is set to $0.5\pi$. As illustrated in Fig.~6(b), the linear resistance maintains a value of zero over a relatively broad energy range of the MBSs, extending from $\varepsilon_m=0$ to $0.5$. As mentioned above, under this specific conditions, i.e., $\phi\neq n\pi$ and $V_2=e^{i\phi}V_1$, the whole system behaves equivalently like a Fano system, consisting of a non-interacting QD connected to two normal leads and another non-interacting QD. Therefore, we redefine the total linear capacitance of the Fano system as:
\begin{equation}
C_{d1}+C_{d2} = -\frac{\dot{\rho}^{0}_{d}(\theta) + \dot{\rho}^{0}_{f}(\theta)}{\dot{\varepsilon}_{ac}(\theta)}\bigg |_{\varepsilon_{ac}\rightarrow 0},
\label{totalCd}
\end{equation}
and we plot this total capacitance as a function of $\varepsilon_d$ in Fig.~7. It is observed that the total capacitance is exactly equal to zero at the case of MBS zero mode, signifying that the normal leads exert no influence on the RC circuit under this condition.
Consequently, when the RC circuit composed of the QD and MBSs becomes decoupled from the leads at zero mode, the circuit's driving mechanism stems from the gate voltage that dynamically modulates the quantum dot's energy level, where periodic voltage variations induce corresponding quantum dot energy oscillations, establishing this as the fundamental excitation source for RC circuit operation - a configuration that directly explains the observed zero linear resistance in Fig.~6(b) and zero linear conductance in Fig.~2(b).
In Fig. 6(b), we find that at the specific phase difference $\phi=\frac{\pi}{2}$ with $\epsilon_{m}=0$, the system's relaxation resistance completely disappears, while Fig. 4(a) simultaneously shows the conductance vanishing under identical conditions. This seemingly paradoxical observation is reconciled by recognizing the fundamental distinction between these quantities: the conductance represents the DC current response while the relaxation resistance describes AC current dynamics - allowing both parameters to concurrently approach zero without contradiction.
\begin{figure}[h]
\centering
\includegraphics[scale=0.21]{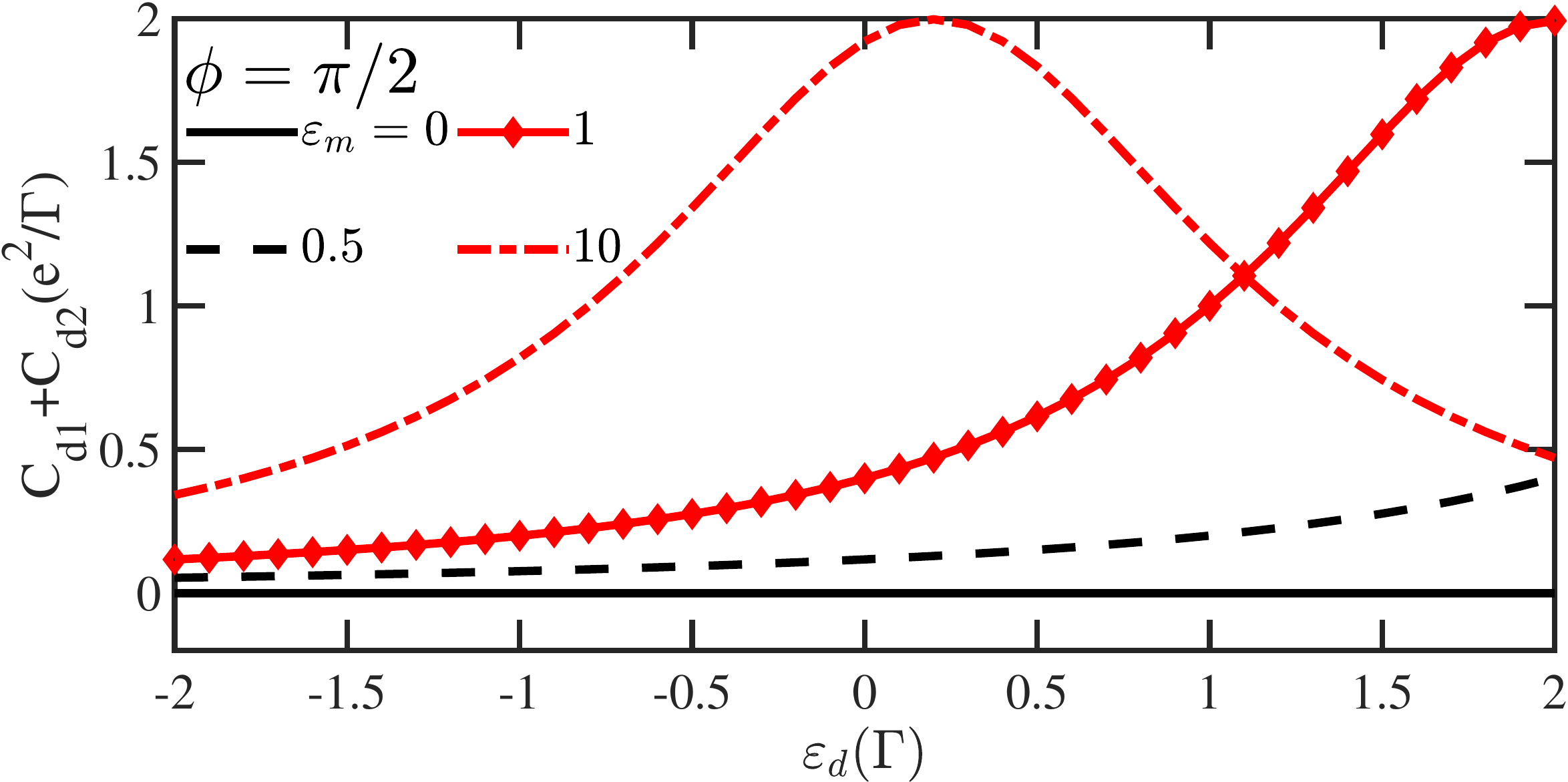}
\caption{(Colour online) The total linear capacitance, defined in Eq.~\eqref{totalCd}, of the QD and MBSs as a function of the dot level for various MBSs modes at $\phi=\pi/2$.}
\end{figure}
\begin{figure}[h]
\centering
\includegraphics[scale=0.21]{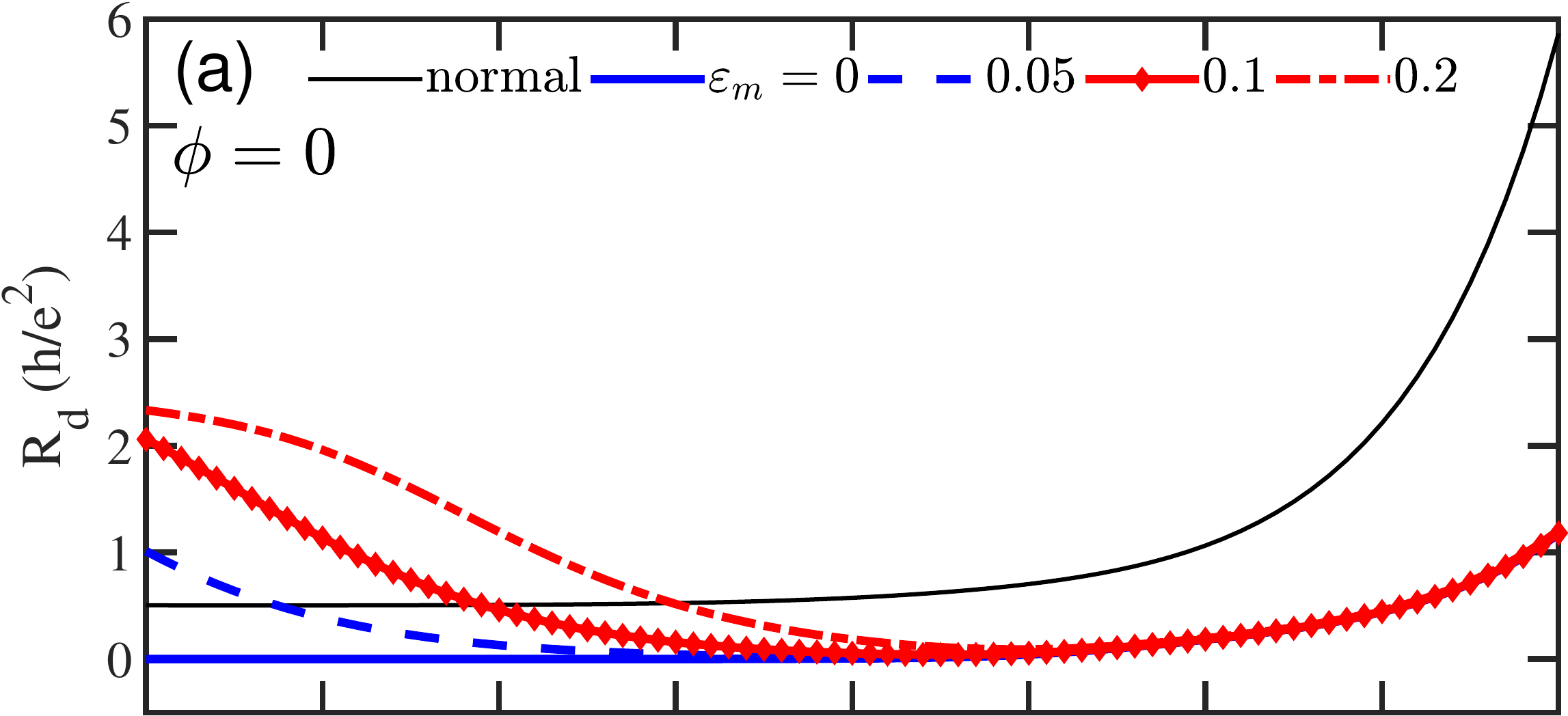}
\includegraphics[scale=0.21]{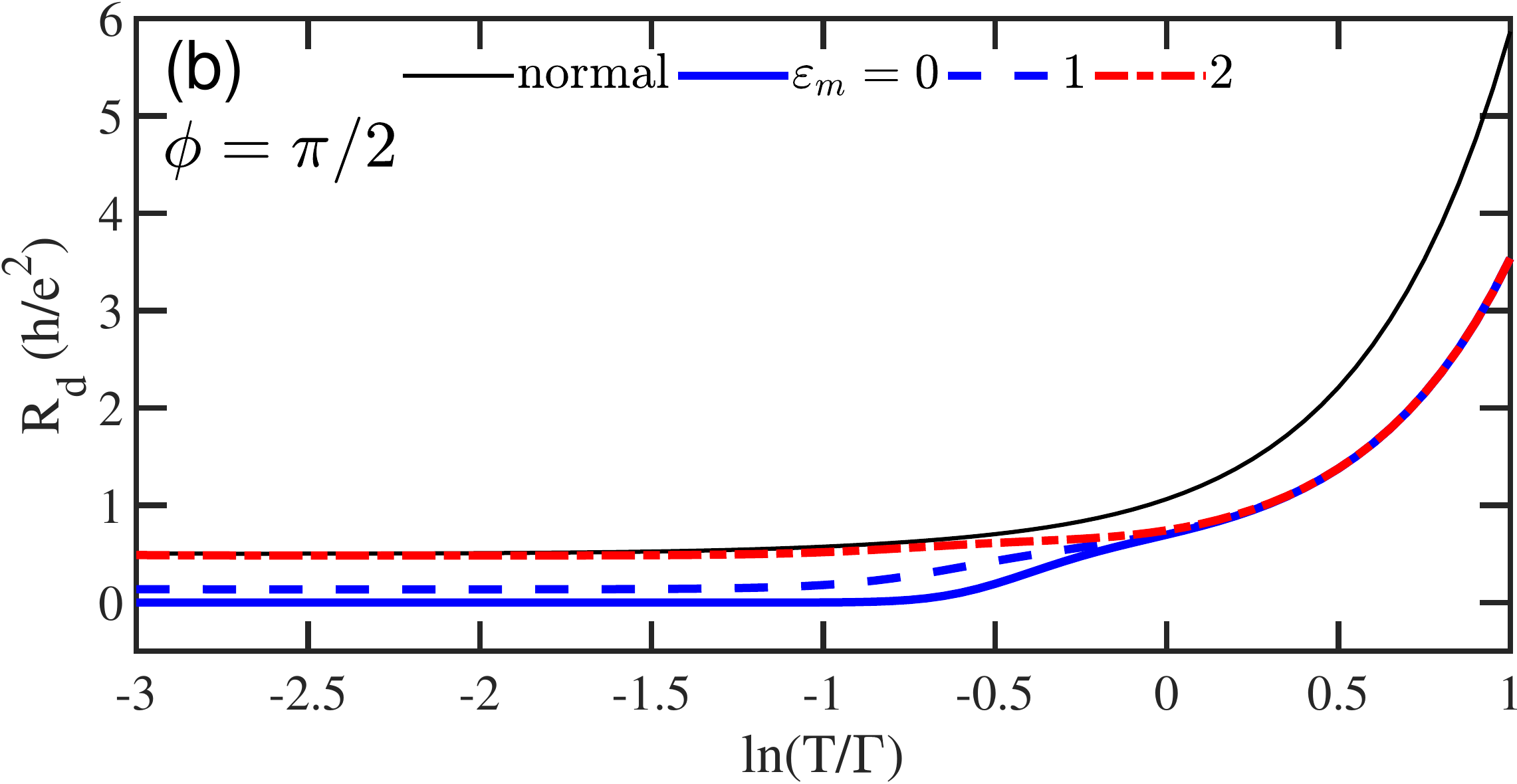}
\caption{(Colour online) The linear relaxation resistance as a function of temperature for various modes $\varepsilon_m$ of the MBSs with $\varepsilon_d = 0$ at different phase differences $\phi=0$ (a) and $\pi/2$ (b), respectively.}
\end{figure}

In our final analysis, we investigate the impact of temperature on the linear relaxation resistance, in Fig.~8, for the particular system with $\varepsilon_d = 0$.
At lower temperatures, the linear resistance is observed to be zero when the MBSs is in the zero mode with $\phi=0$. As the value of $\varepsilon_m$ increases, the resistance also rises, indicative of the system's sensitivity to the MBSs mode. Conversely, at higher temperatures, the resistance decreases, trending towards the value seen at $\varepsilon_m = 0$. Figure 8(b) presents the linear resistance for the situation with $\phi=0.5\pi$. In this instance, the resistance exhibits a monotonic increase with temperature, a behavior attributed to the dissociation of electron-hole pairs.

\section{Conclusion}

In this study, we conduct a theoretical examination of a spinless QD that is linked to two normal leads and two MBSs. By employing the NGF method and Pad$\acute{e}$ expansion, we formulate the EOMs and utilize them to ascertain the linear conductance, linear capacitance, and linear relaxation resistance of the system.

We find that when the phase difference between the MBSs is not an integer multiple of $\pi$ and the MBSs are in the zero mode, the linear conductance is fully suppressed. In this scenario, the conductance remains unresponsive to the MBSs mode. Furthermore, when $\varepsilon_d$ = 0 or $\varepsilon_m$ = 0, the conductance displays a periodicity of $\pi$ as a function of the phase difference; in all other cases, the periodicity is $2\pi$.

Our calculations reveal that in the presence of the zero mode of MBSs, this maximum capacitance value reduces, and the linear capacitance exhibits an increasing trend with the MBSs mode when the phase difference is zero. The linear relaxation resistance displays intriguing behavior: it becomes negligible when the MBSs are in the zero mode, yet it escalates with the MBSs mode. When the phase difference is an integer multiple of $\pi$ the resistance is sensitive to the MBSs mode; conversely, it remains unresponsive under other conditions. This distinction arises due to the distinct mechanisms leading to zero resistance in these two scenarios.

Finally, when the phase difference between two MBSs is zero, we observe that at low temperatures and in the presence of non-zero MBSs modes, the linear resistance is non-zero. As the temperature rises, the resistance decreases, ultimately converging to the value observed when $\varepsilon_m$ = 0.
\bibliography{ref}

\begin{thebibliography}{50}%
\makeatletter
\providecommand \@ifxundefined [1]{%
 \@ifx{#1\undefined}
}%
\providecommand \@ifnum [1]{%
 \ifnum #1\expandafter \@firstoftwo
 \else \expandafter \@secondoftwo
 \fi
}%
\providecommand \@ifx [1]{%
 \ifx #1\expandafter \@firstoftwo
 \else \expandafter \@secondoftwo
 \fi
}%
\providecommand \natexlab [1]{#1}%
\providecommand \enquote  [1]{``#1''}%
\providecommand \bibnamefont  [1]{#1}%
\providecommand \bibfnamefont [1]{#1}%
\providecommand \citenamefont [1]{#1}%
\providecommand \href@noop [0]{\@secondoftwo}%
\providecommand \href [0]{\begingroup \@sanitize@url \@href}%
\providecommand \@href[1]{\@@startlink{#1}\@@href}%
\providecommand \@@href[1]{\endgroup#1\@@endlink}%
\providecommand \@sanitize@url [0]{\catcode `\\12\catcode `\$12\catcode
  `\&12\catcode `\#12\catcode `\^12\catcode `\_12\catcode `\%12\relax}%
\providecommand \@@startlink[1]{}%
\providecommand \@@endlink[0]{}%
\providecommand \url  [0]{\begingroup\@sanitize@url \@url }%
\providecommand \@url [1]{\endgroup\@href {#1}{\urlprefix }}%
\providecommand \urlprefix  [0]{URL }%
\providecommand \Eprint [0]{\href }%
\providecommand \doibase [0]{https://doi.org/}%
\providecommand \selectlanguage [0]{\@gobble}%
\providecommand \bibinfo  [0]{\@secondoftwo}%
\providecommand \bibfield  [0]{\@secondoftwo}%
\providecommand \translation [1]{[#1]}%
\providecommand \BibitemOpen [0]{}%
\providecommand \bibitemStop [0]{}%
\providecommand \bibitemNoStop [0]{.\EOS\space}%
\providecommand \EOS [0]{\spacefactor3000\relax}%
\providecommand \BibitemShut  [1]{\csname bibitem#1\endcsname}%
\let\auto@bib@innerbib\@empty
\bibitem [{\citenamefont {Albrecht}\ \emph {et~al.}(2016)\citenamefont
  {Albrecht}, \citenamefont {Higginbotham}, \citenamefont {Madsen},
  \citenamefont {Kuemmeth}, \citenamefont {Jespersen}, \citenamefont {Nygard},
  \citenamefont {Krogstrup},\ and\ \citenamefont {Marcus}}]{RN24}%
  \BibitemOpen
  \bibfield  {author} {\bibinfo {author} {\bibfnamefont {S.~M.}\ \bibnamefont
  {Albrecht}}, \bibinfo {author} {\bibfnamefont {A.~P.}\ \bibnamefont
  {Higginbotham}}, \bibinfo {author} {\bibfnamefont {M.}~\bibnamefont
  {Madsen}}, \bibinfo {author} {\bibfnamefont {F.}~\bibnamefont {Kuemmeth}},
  \bibinfo {author} {\bibfnamefont {T.~S.}\ \bibnamefont {Jespersen}}, \bibinfo
  {author} {\bibfnamefont {J.}~\bibnamefont {Nygard}}, \bibinfo {author}
  {\bibfnamefont {P.}~\bibnamefont {Krogstrup}},\ and\ \bibinfo {author}
  {\bibfnamefont {C.~M.}\ \bibnamefont {Marcus}},\ }\bibfield  {title}
  {\bibinfo {title} {Exponential protection of zero modes in majorana
  islands},\ }\href {https://doi.org/nature17162} {\bibfield  {journal}
  {\bibinfo  {journal} {Nature}\ }\textbf {\bibinfo {volume} {531}},\ \bibinfo
  {pages} {206} (\bibinfo {year} {2016})}\BibitemShut {NoStop}%
\bibitem [{\citenamefont {Zhang}\ \emph
  {et~al.}(2018{\natexlab{a}})\citenamefont {Zhang}, \citenamefont {Liu},
  \citenamefont {Gazibegovic}, \citenamefont {Xu}, \citenamefont {Logan},
  \citenamefont {Wang}, \citenamefont {van Loo}, \citenamefont {Bommer},
  \citenamefont {de~Moor}, \citenamefont {Car}, \citenamefont {Op~Het~Veld},
  \citenamefont {van Veldhoven}, \citenamefont {Koelling}, \citenamefont
  {Verheijen}, \citenamefont {Pendharkar}, \citenamefont {Pennachio},
  \citenamefont {Shojaei}, \citenamefont {Lee}, \citenamefont {Palmstrom},
  \citenamefont {Bakkers}, \citenamefont {Sarma},\ and\ \citenamefont
  {Kouwenhoven}}]{RN25}%
  \BibitemOpen
  \bibfield  {author} {\bibinfo {author} {\bibfnamefont {H.}~\bibnamefont
  {Zhang}}, \bibinfo {author} {\bibfnamefont {C.~X.}\ \bibnamefont {Liu}},
  \bibinfo {author} {\bibfnamefont {S.}~\bibnamefont {Gazibegovic}}, \bibinfo
  {author} {\bibfnamefont {D.}~\bibnamefont {Xu}}, \bibinfo {author}
  {\bibfnamefont {J.~A.}\ \bibnamefont {Logan}}, \bibinfo {author}
  {\bibfnamefont {G.}~\bibnamefont {Wang}}, \bibinfo {author} {\bibfnamefont
  {N.}~\bibnamefont {van Loo}}, \bibinfo {author} {\bibfnamefont {J.~D.~S.}\
  \bibnamefont {Bommer}}, \bibinfo {author} {\bibfnamefont {M.~W.~A.}\
  \bibnamefont {de~Moor}}, \bibinfo {author} {\bibfnamefont {D.}~\bibnamefont
  {Car}}, \bibinfo {author} {\bibfnamefont {R.~L.~M.}\ \bibnamefont
  {Op~Het~Veld}}, \bibinfo {author} {\bibfnamefont {P.~J.}\ \bibnamefont {van
  Veldhoven}}, \bibinfo {author} {\bibfnamefont {S.}~\bibnamefont {Koelling}},
  \bibinfo {author} {\bibfnamefont {M.~A.}\ \bibnamefont {Verheijen}}, \bibinfo
  {author} {\bibfnamefont {M.}~\bibnamefont {Pendharkar}}, \bibinfo {author}
  {\bibfnamefont {D.~J.}\ \bibnamefont {Pennachio}}, \bibinfo {author}
  {\bibfnamefont {B.}~\bibnamefont {Shojaei}}, \bibinfo {author} {\bibfnamefont
  {J.~S.}\ \bibnamefont {Lee}}, \bibinfo {author} {\bibfnamefont {C.~J.}\
  \bibnamefont {Palmstrom}}, \bibinfo {author} {\bibfnamefont {E.}~\bibnamefont
  {Bakkers}}, \bibinfo {author} {\bibfnamefont {S.~D.}\ \bibnamefont {Sarma}},\
  and\ \bibinfo {author} {\bibfnamefont {L.~P.}\ \bibnamefont {Kouwenhoven}},\
  }\bibfield  {title} {\bibinfo {title} {Quantized majorana conductance},\
  }\href {https://doi.org/nature26142} {\bibfield  {journal} {\bibinfo
  {journal} {Nature}\ }\textbf {\bibinfo {volume} {556}},\ \bibinfo {pages}
  {74} (\bibinfo {year} {2018}{\natexlab{a}})}\BibitemShut {NoStop}%
\bibitem [{\citenamefont {Zhang}\ \emph
  {et~al.}(2018{\natexlab{b}})\citenamefont {Zhang}, \citenamefont {Peng},
  \citenamefont {Xie}, \citenamefont {Li}, \citenamefont {Li}, \citenamefont
  {Yang}, \citenamefont {Guo}, \citenamefont {Qin}, \citenamefont {Chen},
  \citenamefont {Gao}, \citenamefont {Zheng}, \citenamefont {Xiao},\ and\
  \citenamefont {Jia}}]{RN105}%
  \BibitemOpen
  \bibfield  {author} {\bibinfo {author} {\bibfnamefont {G.-F.}\ \bibnamefont
  {Zhang}}, \bibinfo {author} {\bibfnamefont {Y.-G.}\ \bibnamefont {Peng}},
  \bibinfo {author} {\bibfnamefont {H.-Q.}\ \bibnamefont {Xie}}, \bibinfo
  {author} {\bibfnamefont {B.}~\bibnamefont {Li}}, \bibinfo {author}
  {\bibfnamefont {Z.-J.}\ \bibnamefont {Li}}, \bibinfo {author} {\bibfnamefont
  {C.-G.}\ \bibnamefont {Yang}}, \bibinfo {author} {\bibfnamefont {W.-L.}\
  \bibnamefont {Guo}}, \bibinfo {author} {\bibfnamefont {C.-B.}\ \bibnamefont
  {Qin}}, \bibinfo {author} {\bibfnamefont {R.-Y.}\ \bibnamefont {Chen}},
  \bibinfo {author} {\bibfnamefont {Y.}~\bibnamefont {Gao}}, \bibinfo {author}
  {\bibfnamefont {Y.-J.}\ \bibnamefont {Zheng}}, \bibinfo {author}
  {\bibfnamefont {L.-T.}\ \bibnamefont {Xiao}},\ and\ \bibinfo {author}
  {\bibfnamefont {S.-T.}\ \bibnamefont {Jia}},\ }\bibfield  {title} {\bibinfo
  {title} {Linear dipole behavior of single quantum dots encased in metal oxide
  semiconductor nanoparticles films},\ }\href
  {https://doi.org/https://doi.org/10.1007/s11467-019-0916-1} {\bibfield
  {journal} {\bibinfo  {journal} {Frontiers of Physics}\ }\textbf {\bibinfo
  {volume} {14}},\ \bibinfo {pages} {23605} (\bibinfo {year}
  {2018}{\natexlab{b}})}\BibitemShut {NoStop}%
\bibitem [{\citenamefont {Zhang}\ \emph {et~al.}(2019)\citenamefont {Zhang},
  \citenamefont {Yang}, \citenamefont {Ge}, \citenamefont {Peng}, \citenamefont
  {Chen}, \citenamefont {Qin}, \citenamefont {Gao}, \citenamefont {Zhang},
  \citenamefont {Zhong}, \citenamefont {Zheng}, \citenamefont {Xiao},\ and\
  \citenamefont {Jia}}]{RN108}%
  \BibitemOpen
  \bibfield  {author} {\bibinfo {author} {\bibfnamefont {G.-F.}\ \bibnamefont
  {Zhang}}, \bibinfo {author} {\bibfnamefont {C.-G.}\ \bibnamefont {Yang}},
  \bibinfo {author} {\bibfnamefont {Y.}~\bibnamefont {Ge}}, \bibinfo {author}
  {\bibfnamefont {Y.-G.}\ \bibnamefont {Peng}}, \bibinfo {author}
  {\bibfnamefont {R.-Y.}\ \bibnamefont {Chen}}, \bibinfo {author}
  {\bibfnamefont {C.-B.}\ \bibnamefont {Qin}}, \bibinfo {author} {\bibfnamefont
  {Y.}~\bibnamefont {Gao}}, \bibinfo {author} {\bibfnamefont {L.}~\bibnamefont
  {Zhang}}, \bibinfo {author} {\bibfnamefont {H.-Z.}\ \bibnamefont {Zhong}},
  \bibinfo {author} {\bibfnamefont {Y.-J.}\ \bibnamefont {Zheng}}, \bibinfo
  {author} {\bibfnamefont {L.-T.}\ \bibnamefont {Xiao}},\ and\ \bibinfo
  {author} {\bibfnamefont {S.-T.}\ \bibnamefont {Jia}},\ }\bibfield  {title}
  {\bibinfo {title} {Influence of surface charges on the emission polarization
  properties of single cdse/cds dot-in-rods},\ }\href@noop {} {\bibfield
  {journal} {\bibinfo  {journal} {Frontiers of Physics}\ }\textbf {\bibinfo
  {volume} {14}},\ \bibinfo {pages} {63601} (\bibinfo {year}
  {2019})}\BibitemShut {NoStop}%
\bibitem [{\citenamefont {Rokhinson}\ \emph {et~al.}(2012)\citenamefont
  {Rokhinson}, \citenamefont {Liu},\ and\ \citenamefont {Furdyna}}]{RN131}%
  \BibitemOpen
  \bibfield  {author} {\bibinfo {author} {\bibfnamefont {L.~P.}\ \bibnamefont
  {Rokhinson}}, \bibinfo {author} {\bibfnamefont {X.}~\bibnamefont {Liu}},\
  and\ \bibinfo {author} {\bibfnamefont {J.~K.}\ \bibnamefont {Furdyna}},\
  }\bibfield  {title} {\bibinfo {title} {The fractional a.c. josephson effect
  in a semiconductor csuperconductor nanowire as a signature of majorana
  particles},\ }\href {https://doi.org/10.1038/nphys2429} {\bibfield  {journal}
  {\bibinfo  {journal} {Nature Physics}\ }\textbf {\bibinfo {volume} {8}},\
  \bibinfo {pages} {795} (\bibinfo {year} {2012})}\BibitemShut {NoStop}%
\bibitem [{\citenamefont {Das}\ \emph {et~al.}(2012)\citenamefont {Das},
  \citenamefont {Ronen}, \citenamefont {Most}, \citenamefont {Oreg},
  \citenamefont {Heiblum},\ and\ \citenamefont {Shtrikman}}]{RN132}%
  \BibitemOpen
  \bibfield  {author} {\bibinfo {author} {\bibfnamefont {A.}~\bibnamefont
  {Das}}, \bibinfo {author} {\bibfnamefont {Y.}~\bibnamefont {Ronen}}, \bibinfo
  {author} {\bibfnamefont {Y.}~\bibnamefont {Most}}, \bibinfo {author}
  {\bibfnamefont {Y.}~\bibnamefont {Oreg}}, \bibinfo {author} {\bibfnamefont
  {M.}~\bibnamefont {Heiblum}},\ and\ \bibinfo {author} {\bibfnamefont
  {H.}~\bibnamefont {Shtrikman}},\ }\bibfield  {title} {\bibinfo {title}
  {Zero-bias peaks and splitting in an al cinas nanowire topological
  superconductor as a signature of majorana fermions},\ }\href
  {https://doi.org/10.1038/nphys2479} {\bibfield  {journal} {\bibinfo
  {journal} {Nature Physics}\ }\textbf {\bibinfo {volume} {8}},\ \bibinfo
  {pages} {887} (\bibinfo {year} {2012})}\BibitemShut {NoStop}%
\bibitem [{\citenamefont {Churchill}\ \emph {et~al.}(2013)\citenamefont
  {Churchill}, \citenamefont {Fatemi}, \citenamefont {Grove-Rasmussen},
  \citenamefont {Deng}, \citenamefont {Caroff}, \citenamefont {Xu},\ and\
  \citenamefont {Marcus}}]{RN134}%
  \BibitemOpen
  \bibfield  {author} {\bibinfo {author} {\bibfnamefont {H.~O.~H.}\
  \bibnamefont {Churchill}}, \bibinfo {author} {\bibfnamefont {V.}~\bibnamefont
  {Fatemi}}, \bibinfo {author} {\bibfnamefont {K.}~\bibnamefont
  {Grove-Rasmussen}}, \bibinfo {author} {\bibfnamefont {M.~T.}\ \bibnamefont
  {Deng}}, \bibinfo {author} {\bibfnamefont {P.}~\bibnamefont {Caroff}},
  \bibinfo {author} {\bibfnamefont {H.~Q.}\ \bibnamefont {Xu}},\ and\ \bibinfo
  {author} {\bibfnamefont {C.~M.}\ \bibnamefont {Marcus}},\ }\bibfield  {title}
  {\bibinfo {title} {Superconductor-nanowire devices from tunneling to the
  multichannel regime: Zero-bias oscillations and magnetoconductance
  crossover},\ }\href {https://doi.org/10.1103/PhysRevB.87.241401} {\bibfield
  {journal} {\bibinfo  {journal} {Phys. Rev. B}\ }\textbf {\bibinfo {volume}
  {87}},\ \bibinfo {pages} {241401} (\bibinfo {year} {2013})}\BibitemShut
  {NoStop}%
\bibitem [{\citenamefont {Nichele}\ \emph {et~al.}(2017)\citenamefont
  {Nichele}, \citenamefont {Drachmann}, \citenamefont {Whiticar}, \citenamefont
  {O'Farrell}, \citenamefont {Suominen}, \citenamefont {Fornieri},
  \citenamefont {Wang}, \citenamefont {Gardner}, \citenamefont {Thomas},
  \citenamefont {Hatke}, \citenamefont {Krogstrup}, \citenamefont {Manfra},
  \citenamefont {Flensberg},\ and\ \citenamefont {Marcus}}]{RN135}%
  \BibitemOpen
  \bibfield  {author} {\bibinfo {author} {\bibfnamefont {F.}~\bibnamefont
  {Nichele}}, \bibinfo {author} {\bibfnamefont {A.~C.~C.}\ \bibnamefont
  {Drachmann}}, \bibinfo {author} {\bibfnamefont {A.~M.}\ \bibnamefont
  {Whiticar}}, \bibinfo {author} {\bibfnamefont {E.~C.~T.}\ \bibnamefont
  {O'Farrell}}, \bibinfo {author} {\bibfnamefont {H.~J.}\ \bibnamefont
  {Suominen}}, \bibinfo {author} {\bibfnamefont {A.}~\bibnamefont {Fornieri}},
  \bibinfo {author} {\bibfnamefont {T.}~\bibnamefont {Wang}}, \bibinfo {author}
  {\bibfnamefont {G.~C.}\ \bibnamefont {Gardner}}, \bibinfo {author}
  {\bibfnamefont {C.}~\bibnamefont {Thomas}}, \bibinfo {author} {\bibfnamefont
  {A.~T.}\ \bibnamefont {Hatke}}, \bibinfo {author} {\bibfnamefont
  {P.}~\bibnamefont {Krogstrup}}, \bibinfo {author} {\bibfnamefont {M.~J.}\
  \bibnamefont {Manfra}}, \bibinfo {author} {\bibfnamefont {K.}~\bibnamefont
  {Flensberg}},\ and\ \bibinfo {author} {\bibfnamefont {C.~M.}\ \bibnamefont
  {Marcus}},\ }\bibfield  {title} {\bibinfo {title} {Scaling of majorana
  zero-bias conductance peaks},\ }\href
  {https://doi.org/10.1103/PhysRevLett.119.136803} {\bibfield  {journal}
  {\bibinfo  {journal} {Phys Rev Lett}\ }\textbf {\bibinfo {volume} {119}},\
  \bibinfo {pages} {136803} (\bibinfo {year} {2017})}\BibitemShut {NoStop}%
\bibitem [{\citenamefont {Suominen}\ \emph {et~al.}(2017)\citenamefont
  {Suominen}, \citenamefont {Kjaergaard}, \citenamefont {Hamilton},
  \citenamefont {Shabani}, \citenamefont {Palmstrom}, \citenamefont {Marcus},\
  and\ \citenamefont {Nichele}}]{RN136}%
  \BibitemOpen
  \bibfield  {author} {\bibinfo {author} {\bibfnamefont {H.~J.}\ \bibnamefont
  {Suominen}}, \bibinfo {author} {\bibfnamefont {M.}~\bibnamefont
  {Kjaergaard}}, \bibinfo {author} {\bibfnamefont {A.~R.}\ \bibnamefont
  {Hamilton}}, \bibinfo {author} {\bibfnamefont {J.}~\bibnamefont {Shabani}},
  \bibinfo {author} {\bibfnamefont {C.~J.}\ \bibnamefont {Palmstrom}}, \bibinfo
  {author} {\bibfnamefont {C.~M.}\ \bibnamefont {Marcus}},\ and\ \bibinfo
  {author} {\bibfnamefont {F.}~\bibnamefont {Nichele}},\ }\bibfield  {title}
  {\bibinfo {title} {Zero-energy modes from coalescing andreev states in a
  two-dimensional semiconductor-superconductor hybrid platform},\ }\href
  {https://doi.org/10.1103/PhysRevLett.119.176805} {\bibfield  {journal}
  {\bibinfo  {journal} {Phys Rev Lett}\ }\textbf {\bibinfo {volume} {119}},\
  \bibinfo {pages} {176805} (\bibinfo {year} {2017})}\BibitemShut {NoStop}%
\bibitem [{\citenamefont {Eriksson}\ \emph {et~al.}(2015)\citenamefont
  {Eriksson}, \citenamefont {Zazunov}, \citenamefont {Sodano},\ and\
  \citenamefont {Egger}}]{RN40}%
  \BibitemOpen
  \bibfield  {author} {\bibinfo {author} {\bibfnamefont {E.}~\bibnamefont
  {Eriksson}}, \bibinfo {author} {\bibfnamefont {A.}~\bibnamefont {Zazunov}},
  \bibinfo {author} {\bibfnamefont {P.}~\bibnamefont {Sodano}},\ and\ \bibinfo
  {author} {\bibfnamefont {R.}~\bibnamefont {Egger}},\ }\bibfield  {title}
  {\bibinfo {title} {Two-impurity helical majorana problem},\ }\href
  {https://doi.org/10.1103/PhysRevB.91.064501} {\bibfield  {journal} {\bibinfo
  {journal} {Phys. Rev. B}\ }\textbf {\bibinfo {volume} {91}},\ \bibinfo
  {pages} {064501} (\bibinfo {year} {2015})}\BibitemShut {NoStop}%
\bibitem [{\citenamefont {Gao}\ and\ \citenamefont {Gong}(2016)}]{RN56}%
  \BibitemOpen
  \bibfield  {author} {\bibinfo {author} {\bibfnamefont {Z.}~\bibnamefont
  {Gao}}\ and\ \bibinfo {author} {\bibfnamefont {W.-J.}\ \bibnamefont {Gong}},\
  }\bibfield  {title} {\bibinfo {title} {Kondo effect modified by majorana
  doublet at end of a diii-class topological superconductor},\ }\href
  {https://doi.org/10.1103/PhysRevB.94.104506} {\bibfield  {journal} {\bibinfo
  {journal} {Phys. Rev. B}\ }\textbf {\bibinfo {volume} {94}},\ \bibinfo
  {pages} {104506} (\bibinfo {year} {2016})}\BibitemShut {NoStop}%
\bibitem [{\citenamefont {Liu}\ \emph {et~al.}(2015{\natexlab{a}})\citenamefont
  {Liu}, \citenamefont {Cheng},\ and\ \citenamefont {Lutchyn}}]{RN41}%
  \BibitemOpen
  \bibfield  {author} {\bibinfo {author} {\bibfnamefont {D.~E.}\ \bibnamefont
  {Liu}}, \bibinfo {author} {\bibfnamefont {M.}~\bibnamefont {Cheng}},\ and\
  \bibinfo {author} {\bibfnamefont {R.~M.}\ \bibnamefont {Lutchyn}},\
  }\bibfield  {title} {\bibinfo {title} {Probing majorana physics in
  quantum-dot shot-noise experiments},\ }\href
  {https://doi.org/10.1103/PhysRevB.91.081405} {\bibfield  {journal} {\bibinfo
  {journal} {Phys. Rev. B}\ }\textbf {\bibinfo {volume} {91}},\ \bibinfo
  {pages} {081405} (\bibinfo {year} {2015}{\natexlab{a}})}\BibitemShut
  {NoStop}%
\bibitem [{\citenamefont {Liu}\ \emph {et~al.}(2013)\citenamefont {Liu},
  \citenamefont {Levchenko},\ and\ \citenamefont {Baranger}}]{RN74}%
  \BibitemOpen
  \bibfield  {author} {\bibinfo {author} {\bibfnamefont {D.~E.}\ \bibnamefont
  {Liu}}, \bibinfo {author} {\bibfnamefont {A.}~\bibnamefont {Levchenko}},\
  and\ \bibinfo {author} {\bibfnamefont {H.~U.}\ \bibnamefont {Baranger}},\
  }\bibfield  {title} {\bibinfo {title} {Floquet majorana fermions for
  topological qubits in superconducting devices and cold-atom systems},\ }\href
  {https://doi.org/10.1103/PhysRevLett.111.047002} {\bibfield  {journal}
  {\bibinfo  {journal} {Phys Rev Lett}\ }\textbf {\bibinfo {volume} {111}},\
  \bibinfo {pages} {047002} (\bibinfo {year} {2013})}\BibitemShut {NoStop}%
\bibitem [{\citenamefont {Liu}\ \emph {et~al.}(2015{\natexlab{b}})\citenamefont
  {Liu}, \citenamefont {Levchenko},\ and\ \citenamefont {Lutchyn}}]{RN52}%
  \BibitemOpen
  \bibfield  {author} {\bibinfo {author} {\bibfnamefont {D.~E.}\ \bibnamefont
  {Liu}}, \bibinfo {author} {\bibfnamefont {A.}~\bibnamefont {Levchenko}},\
  and\ \bibinfo {author} {\bibfnamefont {R.~M.}\ \bibnamefont {Lutchyn}},\
  }\bibfield  {title} {\bibinfo {title} {Majorana zero modes choose euler
  numbers as revealed by full counting statistics},\ }\href
  {https://doi.org/10.1103/PhysRevB.92.205422} {\bibfield  {journal} {\bibinfo
  {journal} {Phys. Rev. B}\ }\textbf {\bibinfo {volume} {92}},\ \bibinfo
  {pages} {205422} (\bibinfo {year} {2015}{\natexlab{b}})}\BibitemShut
  {NoStop}%
\bibitem [{\citenamefont {L\"u}\ \emph {et~al.}(2012)\citenamefont {L\"u},
  \citenamefont {Lu},\ and\ \citenamefont {Shen}}]{RN45}%
  \BibitemOpen
  \bibfield  {author} {\bibinfo {author} {\bibfnamefont {H.-F.}\ \bibnamefont
  {L\"u}}, \bibinfo {author} {\bibfnamefont {H.-Z.}\ \bibnamefont {Lu}},\ and\
  \bibinfo {author} {\bibfnamefont {S.-Q.}\ \bibnamefont {Shen}},\ }\bibfield
  {title} {\bibinfo {title} {Nonlocal noise cross correlation mediated by
  entangled majorana fermions},\ }\href
  {https://doi.org/10.1103/PhysRevB.86.075318} {\bibfield  {journal} {\bibinfo
  {journal} {Phys. Rev. B}\ }\textbf {\bibinfo {volume} {86}},\ \bibinfo
  {pages} {075318} (\bibinfo {year} {2012})}\BibitemShut {NoStop}%
\bibitem [{\citenamefont {Gong}\ \emph {et~al.}(2014)\citenamefont {Gong},
  \citenamefont {Zhang}, \citenamefont {Li}, \citenamefont {Yi},\ and\
  \citenamefont {Zheng}}]{RN46}%
  \BibitemOpen
  \bibfield  {author} {\bibinfo {author} {\bibfnamefont {W.-J.}\ \bibnamefont
  {Gong}}, \bibinfo {author} {\bibfnamefont {S.-F.}\ \bibnamefont {Zhang}},
  \bibinfo {author} {\bibfnamefont {Z.-C.}\ \bibnamefont {Li}}, \bibinfo
  {author} {\bibfnamefont {G.}~\bibnamefont {Yi}},\ and\ \bibinfo {author}
  {\bibfnamefont {Y.-S.}\ \bibnamefont {Zheng}},\ }\bibfield  {title} {\bibinfo
  {title} {Detection of a majorana fermion zero mode by a t-shaped quantum-dot
  structure},\ }\href {https://doi.org/10.1103/PhysRevB.89.245413} {\bibfield
  {journal} {\bibinfo  {journal} {Phys. Rev. B}\ }\textbf {\bibinfo {volume}
  {89}},\ \bibinfo {pages} {245413} (\bibinfo {year} {2014})}\BibitemShut
  {NoStop}%
\bibitem [{\citenamefont {Tripathi}\ \emph {et~al.}(2019)\citenamefont
  {Tripathi}, \citenamefont {Rao},\ and\ \citenamefont {Das}}]{RN32}%
  \BibitemOpen
  \bibfield  {author} {\bibinfo {author} {\bibfnamefont {K.~M.}\ \bibnamefont
  {Tripathi}}, \bibinfo {author} {\bibfnamefont {S.}~\bibnamefont {Rao}},\ and\
  \bibinfo {author} {\bibfnamefont {S.}~\bibnamefont {Das}},\ }\bibfield
  {title} {\bibinfo {title} {Quantum charge pumping through majorana bound
  states},\ }\href {https://doi.org/10.1103/PhysRevB.99.085435} {\bibfield
  {journal} {\bibinfo  {journal} {Phys. Rev. B}\ }\textbf {\bibinfo {volume}
  {99}},\ \bibinfo {pages} {085435} (\bibinfo {year} {2019})}\BibitemShut
  {NoStop}%
\bibitem [{\citenamefont {Smirnov}(2019)}]{RN72}%
  \BibitemOpen
  \bibfield  {author} {\bibinfo {author} {\bibfnamefont {S.}~\bibnamefont
  {Smirnov}},\ }\bibfield  {title} {\bibinfo {title} {Majorana finite-frequency
  nonequilibrium quantum noise},\ }\href
  {https://doi.org/10.1103/PhysRevB.99.165427} {\bibfield  {journal} {\bibinfo
  {journal} {Phys. Rev. B}\ }\textbf {\bibinfo {volume} {99}},\ \bibinfo
  {pages} {165427} (\bibinfo {year} {2019})}\BibitemShut {NoStop}%
\bibitem [{\citenamefont {Schuray}\ \emph {et~al.}(2017)\citenamefont
  {Schuray}, \citenamefont {Weithofer},\ and\ \citenamefont {Recher}}]{RN66}%
  \BibitemOpen
  \bibfield  {author} {\bibinfo {author} {\bibfnamefont {A.}~\bibnamefont
  {Schuray}}, \bibinfo {author} {\bibfnamefont {L.}~\bibnamefont {Weithofer}},\
  and\ \bibinfo {author} {\bibfnamefont {P.}~\bibnamefont {Recher}},\
  }\bibfield  {title} {\bibinfo {title} {Fano resonances in majorana bound
  states--quantum dot hybrid systems},\ }\href
  {https://doi.org/10.1103/PhysRevB.96.085417} {\bibfield  {journal} {\bibinfo
  {journal} {Phys. Rev. B}\ }\textbf {\bibinfo {volume} {96}},\ \bibinfo
  {pages} {085417} (\bibinfo {year} {2017})}\BibitemShut {NoStop}%
\bibitem [{\citenamefont {Ricco}\ \emph {et~al.}(2016)\citenamefont {Ricco},
  \citenamefont {Marques}, \citenamefont {Dessotti}, \citenamefont {Machado},
  \citenamefont {de~Souza},\ and\ \citenamefont {Seridonio}}]{RN53}%
  \BibitemOpen
  \bibfield  {author} {\bibinfo {author} {\bibfnamefont {L.~S.}\ \bibnamefont
  {Ricco}}, \bibinfo {author} {\bibfnamefont {Y.}~\bibnamefont {Marques}},
  \bibinfo {author} {\bibfnamefont {F.~A.}\ \bibnamefont {Dessotti}}, \bibinfo
  {author} {\bibfnamefont {R.~S.}\ \bibnamefont {Machado}}, \bibinfo {author}
  {\bibfnamefont {M.}~\bibnamefont {de~Souza}},\ and\ \bibinfo {author}
  {\bibfnamefont {A.~C.}\ \bibnamefont {Seridonio}},\ }\bibfield  {title}
  {\bibinfo {title} {Decay of bound states in the continuum of majorana
  fermions induced by vacuum fluctuations: Proposal of qubit technology},\
  }\href {https://doi.org/10.1103/PhysRevB.93.165116} {\bibfield  {journal}
  {\bibinfo  {journal} {Phys. Rev. B}\ }\textbf {\bibinfo {volume} {93}},\
  \bibinfo {pages} {165116} (\bibinfo {year} {2016})}\BibitemShut {NoStop}%
\bibitem [{\citenamefont {Golub}(2015)}]{RN42}%
  \BibitemOpen
  \bibfield  {author} {\bibinfo {author} {\bibfnamefont {A.}~\bibnamefont
  {Golub}},\ }\bibfield  {title} {\bibinfo {title} {Multiple andreev
  reflections in $s$-wave superconductor--quantum dot--topological
  superconductor tunnel junctions and majorana bound states},\ }\href
  {https://doi.org/10.1103/PhysRevB.91.205105} {\bibfield  {journal} {\bibinfo
  {journal} {Phys. Rev. B}\ }\textbf {\bibinfo {volume} {91}},\ \bibinfo
  {pages} {205105} (\bibinfo {year} {2015})}\BibitemShut {NoStop}%
\bibitem [{\citenamefont {Vernek}\ \emph {et~al.}(2014)\citenamefont {Vernek},
  \citenamefont {Penteado}, \citenamefont {Seridonio},\ and\ \citenamefont
  {Egues}}]{RN87}%
  \BibitemOpen
  \bibfield  {author} {\bibinfo {author} {\bibfnamefont {E.}~\bibnamefont
  {Vernek}}, \bibinfo {author} {\bibfnamefont {P.~H.}\ \bibnamefont
  {Penteado}}, \bibinfo {author} {\bibfnamefont {A.~C.}\ \bibnamefont
  {Seridonio}},\ and\ \bibinfo {author} {\bibfnamefont {J.~C.}\ \bibnamefont
  {Egues}},\ }\bibfield  {title} {\bibinfo {title} {Subtle leakage of a
  majorana mode into a quantum dot},\ }\href
  {https://doi.org/10.1103/PhysRevB.89.165314} {\bibfield  {journal} {\bibinfo
  {journal} {Phys. Rev. B}\ }\textbf {\bibinfo {volume} {89}},\ \bibinfo
  {pages} {165314} (\bibinfo {year} {2014})}\BibitemShut {NoStop}%
\bibitem [{\citenamefont {L\'opez}\ \emph {et~al.}(2014)\citenamefont
  {L\'opez}, \citenamefont {Lee}, \citenamefont {Serra},\ and\ \citenamefont
  {Lim}}]{RN28}%
  \BibitemOpen
  \bibfield  {author} {\bibinfo {author} {\bibfnamefont {R.}~\bibnamefont
  {L\'opez}}, \bibinfo {author} {\bibfnamefont {M.}~\bibnamefont {Lee}},
  \bibinfo {author} {\bibfnamefont {L.~m.~c.}\ \bibnamefont {Serra}},\ and\
  \bibinfo {author} {\bibfnamefont {J.~S.}\ \bibnamefont {Lim}},\ }\bibfield
  {title} {\bibinfo {title} {Thermoelectrical detection of majorana states},\
  }\href {https://doi.org/10.1103/PhysRevB.89.205418} {\bibfield  {journal}
  {\bibinfo  {journal} {Phys. Rev. B}\ }\textbf {\bibinfo {volume} {89}},\
  \bibinfo {pages} {205418} (\bibinfo {year} {2014})}\BibitemShut {NoStop}%
\bibitem [{\citenamefont {Karzig}\ \emph {et~al.}(2017)\citenamefont {Karzig},
  \citenamefont {Knapp}, \citenamefont {Lutchyn}, \citenamefont {Bonderson},
  \citenamefont {Hastings}, \citenamefont {Nayak}, \citenamefont {Alicea},
  \citenamefont {Flensberg}, \citenamefont {Plugge}, \citenamefont {Oreg},
  \citenamefont {Marcus},\ and\ \citenamefont {Freedman}}]{RN200}%
  \BibitemOpen
  \bibfield  {author} {\bibinfo {author} {\bibfnamefont {T.}~\bibnamefont
  {Karzig}}, \bibinfo {author} {\bibfnamefont {C.}~\bibnamefont {Knapp}},
  \bibinfo {author} {\bibfnamefont {R.~M.}\ \bibnamefont {Lutchyn}}, \bibinfo
  {author} {\bibfnamefont {P.}~\bibnamefont {Bonderson}}, \bibinfo {author}
  {\bibfnamefont {M.~B.}\ \bibnamefont {Hastings}}, \bibinfo {author}
  {\bibfnamefont {C.}~\bibnamefont {Nayak}}, \bibinfo {author} {\bibfnamefont
  {J.}~\bibnamefont {Alicea}}, \bibinfo {author} {\bibfnamefont
  {K.}~\bibnamefont {Flensberg}}, \bibinfo {author} {\bibfnamefont
  {S.}~\bibnamefont {Plugge}}, \bibinfo {author} {\bibfnamefont
  {Y.}~\bibnamefont {Oreg}}, \bibinfo {author} {\bibfnamefont {C.~M.}\
  \bibnamefont {Marcus}},\ and\ \bibinfo {author} {\bibfnamefont {M.~H.}\
  \bibnamefont {Freedman}},\ }\bibfield  {title} {\bibinfo {title} {Scalable
  designs for quasiparticle-poisoning-protected topological quantum computation
  with majorana zero modes},\ }\href
  {https://doi.org/10.1103/PhysRevB.95.235305} {\bibfield  {journal} {\bibinfo
  {journal} {Phys. Rev. B}\ }\textbf {\bibinfo {volume} {95}},\ \bibinfo
  {pages} {235305} (\bibinfo {year} {2017})}\BibitemShut {NoStop}%
\bibitem [{\citenamefont {Hyart}\ \emph {et~al.}(2013)\citenamefont {Hyart},
  \citenamefont {van Heck}, \citenamefont {Fulga}, \citenamefont {Burrello},
  \citenamefont {Akhmerov},\ and\ \citenamefont {Beenakker}}]{RN201}%
  \BibitemOpen
  \bibfield  {author} {\bibinfo {author} {\bibfnamefont {T.}~\bibnamefont
  {Hyart}}, \bibinfo {author} {\bibfnamefont {B.}~\bibnamefont {van Heck}},
  \bibinfo {author} {\bibfnamefont {I.~C.}\ \bibnamefont {Fulga}}, \bibinfo
  {author} {\bibfnamefont {M.}~\bibnamefont {Burrello}}, \bibinfo {author}
  {\bibfnamefont {A.~R.}\ \bibnamefont {Akhmerov}},\ and\ \bibinfo {author}
  {\bibfnamefont {C.~W.~J.}\ \bibnamefont {Beenakker}},\ }\bibfield  {title}
  {\bibinfo {title} {Flux-controlled quantum computation with majorana
  fermions},\ }\href {https://doi.org/10.1103/PhysRevB.88.035121} {\bibfield
  {journal} {\bibinfo  {journal} {Phys. Rev. B}\ }\textbf {\bibinfo {volume}
  {88}},\ \bibinfo {pages} {035121} (\bibinfo {year} {2013})}\BibitemShut
  {NoStop}%
\bibitem [{\citenamefont {Hoffman}\ \emph {et~al.}(2016)\citenamefont
  {Hoffman}, \citenamefont {Schrade}, \citenamefont {Klinovaja},\ and\
  \citenamefont {Loss}}]{RN202}%
  \BibitemOpen
  \bibfield  {author} {\bibinfo {author} {\bibfnamefont {S.}~\bibnamefont
  {Hoffman}}, \bibinfo {author} {\bibfnamefont {C.}~\bibnamefont {Schrade}},
  \bibinfo {author} {\bibfnamefont {J.}~\bibnamefont {Klinovaja}},\ and\
  \bibinfo {author} {\bibfnamefont {D.}~\bibnamefont {Loss}},\ }\bibfield
  {title} {\bibinfo {title} {Universal quantum computation with hybrid
  spin-majorana qubits},\ }\href {https://doi.org/10.1103/PhysRevB.94.045316}
  {\bibfield  {journal} {\bibinfo  {journal} {Phys. Rev. B}\ }\textbf {\bibinfo
  {volume} {94}},\ \bibinfo {pages} {045316} (\bibinfo {year}
  {2016})}\BibitemShut {NoStop}%
\bibitem [{\citenamefont {Liu}\ and\ \citenamefont {Baranger}(2011)}]{RN27}%
  \BibitemOpen
  \bibfield  {author} {\bibinfo {author} {\bibfnamefont {D.~E.}\ \bibnamefont
  {Liu}}\ and\ \bibinfo {author} {\bibfnamefont {H.~U.}\ \bibnamefont
  {Baranger}},\ }\bibfield  {title} {\bibinfo {title} {Detecting a
  majorana-fermion zero mode using a quantum dot},\ }\href
  {https://doi.org/10.1103/PhysRevB.84.201308} {\bibfield  {journal} {\bibinfo
  {journal} {Phys. Rev. B}\ }\textbf {\bibinfo {volume} {84}},\ \bibinfo
  {pages} {201308} (\bibinfo {year} {2011})}\BibitemShut {NoStop}%
\bibitem [{\citenamefont {Dessotti}\ \emph {et~al.}(2016)\citenamefont
  {Dessotti}, \citenamefont {Ricco}, \citenamefont {Marques}, \citenamefont
  {Guessi}, \citenamefont {Yoshida}, \citenamefont {Figueira}, \citenamefont
  {de~Souza}, \citenamefont {Sodano},\ and\ \citenamefont {Seridonio}}]{RN57}%
  \BibitemOpen
  \bibfield  {author} {\bibinfo {author} {\bibfnamefont {F.~A.}\ \bibnamefont
  {Dessotti}}, \bibinfo {author} {\bibfnamefont {L.~S.}\ \bibnamefont {Ricco}},
  \bibinfo {author} {\bibfnamefont {Y.}~\bibnamefont {Marques}}, \bibinfo
  {author} {\bibfnamefont {L.~H.}\ \bibnamefont {Guessi}}, \bibinfo {author}
  {\bibfnamefont {M.}~\bibnamefont {Yoshida}}, \bibinfo {author} {\bibfnamefont
  {M.~S.}\ \bibnamefont {Figueira}}, \bibinfo {author} {\bibfnamefont
  {M.}~\bibnamefont {de~Souza}}, \bibinfo {author} {\bibfnamefont
  {P.}~\bibnamefont {Sodano}},\ and\ \bibinfo {author} {\bibfnamefont {A.~C.}\
  \bibnamefont {Seridonio}},\ }\bibfield  {title} {\bibinfo {title} {Unveiling
  majorana quasiparticles by a quantum phase transition: Proposal of a current
  switch},\ }\href {https://doi.org/10.1103/PhysRevB.94.125426} {\bibfield
  {journal} {\bibinfo  {journal} {Phys. Rev. B}\ }\textbf {\bibinfo {volume}
  {94}},\ \bibinfo {pages} {125426} (\bibinfo {year} {2016})}\BibitemShut
  {NoStop}%
\bibitem [{\citenamefont {Ramos-Andrade}\ \emph {et~al.}(2018)\citenamefont
  {Ramos-Andrade}, \citenamefont {Orellana},\ and\ \citenamefont
  {Ulloa}}]{RN80}%
  \BibitemOpen
  \bibfield  {author} {\bibinfo {author} {\bibfnamefont {J.~P.}\ \bibnamefont
  {Ramos-Andrade}}, \bibinfo {author} {\bibfnamefont {P.~A.}\ \bibnamefont
  {Orellana}},\ and\ \bibinfo {author} {\bibfnamefont {S.~E.}\ \bibnamefont
  {Ulloa}},\ }\bibfield  {title} {\bibinfo {title} {Detecting coupling of
  majorana bound states with an aharonov-bohm interferometer},\ }\href
  {https://doi.org/10.1088/1361-648X/aaa1b2} {\bibfield  {journal} {\bibinfo
  {journal} {J Phys Condens Matter}\ }\textbf {\bibinfo {volume} {30}},\
  \bibinfo {pages} {045301} (\bibinfo {year} {2018})}\BibitemShut {NoStop}%
\bibitem [{\citenamefont {Ricco}\ \emph {et~al.}(2018)\citenamefont {Ricco},
  \citenamefont {Campo}, \citenamefont {Shelykh},\ and\ \citenamefont
  {Seridonio}}]{RN69}%
  \BibitemOpen
  \bibfield  {author} {\bibinfo {author} {\bibfnamefont {L.~S.}\ \bibnamefont
  {Ricco}}, \bibinfo {author} {\bibfnamefont {V.~L.}\ \bibnamefont {Campo}},
  \bibinfo {author} {\bibfnamefont {I.~A.}\ \bibnamefont {Shelykh}},\ and\
  \bibinfo {author} {\bibfnamefont {A.~C.}\ \bibnamefont {Seridonio}},\
  }\bibfield  {title} {\bibinfo {title} {Majorana oscillations modulated by
  fano interference and degree of nonlocality in a topological
  superconducting-nanowire--quantum-dot system},\ }\href
  {https://doi.org/10.1103/PhysRevB.98.075142} {\bibfield  {journal} {\bibinfo
  {journal} {Phys. Rev. B}\ }\textbf {\bibinfo {volume} {98}},\ \bibinfo
  {pages} {075142} (\bibinfo {year} {2018})}\BibitemShut {NoStop}%
\bibitem [{\citenamefont {Wang}\ \emph {et~al.}(2018)\citenamefont {Wang},
  \citenamefont {Zhang}, \citenamefont {Jiang}, \citenamefont {Yi},\ and\
  \citenamefont {Gong}}]{RN7}%
  \BibitemOpen
  \bibfield  {author} {\bibinfo {author} {\bibfnamefont {X.-Q.}\ \bibnamefont
  {Wang}}, \bibinfo {author} {\bibfnamefont {S.-F.}\ \bibnamefont {Zhang}},
  \bibinfo {author} {\bibfnamefont {C.}~\bibnamefont {Jiang}}, \bibinfo
  {author} {\bibfnamefont {G.-Y.}\ \bibnamefont {Yi}},\ and\ \bibinfo {author}
  {\bibfnamefont {W.-J.}\ \bibnamefont {Gong}},\ }\bibfield  {title} {\bibinfo
  {title} {Fano antiresonance realized by majorana bound states coupled to one
  quantum-dot system},\ }\href
  {https://doi.org/https://doi.org/10.1016/j.physe.2018.07.005} {\bibfield
  {journal} {\bibinfo  {journal} {Physica E: Low-dimensional Systems and
  Nanostructures}\ }\textbf {\bibinfo {volume} {104}},\ \bibinfo {pages} {1}
  (\bibinfo {year} {2018})}\BibitemShut {NoStop}%
\bibitem [{\citenamefont {Yang}(2019)}]{RN107}%
  \BibitemOpen
  \bibfield  {author} {\bibinfo {author} {\bibfnamefont {F.-B.}\ \bibnamefont
  {Yang}},\ }\bibfield  {title} {\bibinfo {title} {Aharonov-bohm interferometer
  in a t-shaped quantum dot embedded in majorana bound states},\ }\href
  {https://doi.org/10.1088/0253-6102/71/8/1024} {\bibfield  {journal} {\bibinfo
   {journal} {Communications in Theoretical Physics}\ }\textbf {\bibinfo
  {volume} {71}},\ \bibinfo {pages} {1024} (\bibinfo {year}
  {2019})}\BibitemShut {NoStop}%
\bibitem [{\citenamefont {Lee}\ and\ \citenamefont {Choi}(2014)}]{RN36}%
  \BibitemOpen
  \bibfield  {author} {\bibinfo {author} {\bibfnamefont {M.}~\bibnamefont
  {Lee}}\ and\ \bibinfo {author} {\bibfnamefont {M.~S.}\ \bibnamefont {Choi}},\
  }\bibfield  {title} {\bibinfo {title} {Quantum resistor-capacitor circuit
  with majorana fermion modes in a chiral topological superconductor},\ }\href
  {https://doi.org/10.1103/PhysRevLett.113.076801} {\bibfield  {journal}
  {\bibinfo  {journal} {Phys Rev Lett}\ }\textbf {\bibinfo {volume} {113}},\
  \bibinfo {pages} {076801} (\bibinfo {year} {2014})}\BibitemShut {NoStop}%
\bibitem [{\citenamefont {Lee}(2019)}]{RN8}%
  \BibitemOpen
  \bibfield  {author} {\bibinfo {author} {\bibfnamefont {M.}~\bibnamefont
  {Lee}},\ }\bibfield  {title} {\bibinfo {title} {Effects of majorana bound
  states on dissipation and charging in a quantum resistor-capacitor circuit},\
  }\href {https://doi.org/10.1016/j.cap.2018.08.001} {\bibfield  {journal}
  {\bibinfo  {journal} {Current Applied Physics}\ }\textbf {\bibinfo {volume}
  {19}},\ \bibinfo {pages} {246} (\bibinfo {year} {2019})}\BibitemShut
  {NoStop}%
\bibitem [{\citenamefont {Yang}\ and\ \citenamefont {jiang
  Liu}(2024)}]{twoMBSs1}%
  \BibitemOpen
  \bibfield  {author} {\bibinfo {author} {\bibfnamefont {F.-B.}\ \bibnamefont
  {Yang}}\ and\ \bibinfo {author} {\bibfnamefont {H.}~\bibnamefont {jiang
  Liu}},\ }\bibfield  {title} {\bibinfo {title} {Spin-dependent electron-phonon
  interaction on quantum transport through an aharonov-bohm interferometer with
  a quantum dot connected to majorana zero mode},\ }\href
  {https://doi.org/https://doi.org/10.1016/j.rinp.2024.108065} {\bibfield
  {journal} {\bibinfo  {journal} {Results in Physics}\ }\textbf {\bibinfo
  {volume} {67}},\ \bibinfo {pages} {108065} (\bibinfo {year}
  {2024})}\BibitemShut {NoStop}%
\bibitem [{\citenamefont {Wrześniewski}\ and\ \citenamefont
  {Weymann}(2024)}]{twoMBSs2}%
  \BibitemOpen
  \bibfield  {author} {\bibinfo {author} {\bibfnamefont {K.}~\bibnamefont
  {Wrześniewski}}\ and\ \bibinfo {author} {\bibfnamefont {I.}~\bibnamefont
  {Weymann}},\ }\bibfield  {title} {\bibinfo {title} {Cross-correlations
  between currents and tunnel magnetoresistance in interacting double quantum
  dot-majorana wire system},\ }\href
  {https://doi.org/10.1038/s41598-024-58344-9} {\bibfield  {journal} {\bibinfo
  {journal} {Scientific Reports}\ }\textbf {\bibinfo {volume} {14}},\ \bibinfo
  {pages} {7815} (\bibinfo {year} {2024})}\BibitemShut {NoStop}%
\bibitem [{\citenamefont {Taranko}\ \emph {et~al.}(2024)\citenamefont
  {Taranko}, \citenamefont {Wrze\ifmmode~\acute{s}\else \'{s}\fi{}niewski},
  \citenamefont {Weymann},\ and\ \citenamefont {Doma\ifmmode~\acute{n}\else
  \'{n}\fi{}ski}}]{twoMBSs3}%
  \BibitemOpen
  \bibfield  {author} {\bibinfo {author} {\bibfnamefont {R.}~\bibnamefont
  {Taranko}}, \bibinfo {author} {\bibfnamefont {K.}~\bibnamefont
  {Wrze\ifmmode~\acute{s}\else \'{s}\fi{}niewski}}, \bibinfo {author}
  {\bibfnamefont {I.}~\bibnamefont {Weymann}},\ and\ \bibinfo {author}
  {\bibfnamefont {T.}~\bibnamefont {Doma\ifmmode~\acute{n}\else
  \'{n}\fi{}ski}},\ }\bibfield  {title} {\bibinfo {title} {Transient effects in
  quantum dots contacted via topological superconductor},\ }\href
  {https://doi.org/10.1103/PhysRevB.110.035413} {\bibfield  {journal} {\bibinfo
   {journal} {Phys. Rev. B}\ }\textbf {\bibinfo {volume} {110}},\ \bibinfo
  {pages} {035413} (\bibinfo {year} {2024})}\BibitemShut {NoStop}%
\bibitem [{\citenamefont {Souto}\ \emph {et~al.}(2023)\citenamefont {Souto},
  \citenamefont {Tsintzis}, \citenamefont {Leijnse},\ and\ \citenamefont
  {Danon}}]{twoMBSs4}%
  \BibitemOpen
  \bibfield  {author} {\bibinfo {author} {\bibfnamefont {R.~S.}\ \bibnamefont
  {Souto}}, \bibinfo {author} {\bibfnamefont {A.}~\bibnamefont {Tsintzis}},
  \bibinfo {author} {\bibfnamefont {M.}~\bibnamefont {Leijnse}},\ and\ \bibinfo
  {author} {\bibfnamefont {J.}~\bibnamefont {Danon}},\ }\bibfield  {title}
  {\bibinfo {title} {Probing majorana localization in minimal kitaev chains
  through a quantum dot},\ }\href
  {https://doi.org/10.1103/PhysRevResearch.5.043182} {\bibfield  {journal}
  {\bibinfo  {journal} {Phys. Rev. Res.}\ }\textbf {\bibinfo {volume} {5}},\
  \bibinfo {pages} {043182} (\bibinfo {year} {2023})}\BibitemShut {NoStop}%
\bibitem [{\citenamefont {Zocher}\ and\ \citenamefont
  {Rosenow}(2013)}]{twoMBSs5}%
  \BibitemOpen
  \bibfield  {author} {\bibinfo {author} {\bibfnamefont {B.}~\bibnamefont
  {Zocher}}\ and\ \bibinfo {author} {\bibfnamefont {B.}~\bibnamefont
  {Rosenow}},\ }\bibfield  {title} {\bibinfo {title} {Modulation of
  majorana-induced current cross-correlations by quantum dots},\ }\href
  {https://doi.org/10.1103/PhysRevLett.111.036802} {\bibfield  {journal}
  {\bibinfo  {journal} {Phys. Rev. Lett.}\ }\textbf {\bibinfo {volume} {111}},\
  \bibinfo {pages} {036802} (\bibinfo {year} {2013})}\BibitemShut {NoStop}%
\bibitem [{\citenamefont {Manousakis}\ \emph {et~al.}(2020)\citenamefont
  {Manousakis}, \citenamefont {Wille}, \citenamefont {Altland}, \citenamefont
  {Egger}, \citenamefont {Flensberg},\ and\ \citenamefont
  {Hassler}}]{twoMBSs6}%
  \BibitemOpen
  \bibfield  {author} {\bibinfo {author} {\bibfnamefont {J.}~\bibnamefont
  {Manousakis}}, \bibinfo {author} {\bibfnamefont {C.}~\bibnamefont {Wille}},
  \bibinfo {author} {\bibfnamefont {A.}~\bibnamefont {Altland}}, \bibinfo
  {author} {\bibfnamefont {R.}~\bibnamefont {Egger}}, \bibinfo {author}
  {\bibfnamefont {K.}~\bibnamefont {Flensberg}},\ and\ \bibinfo {author}
  {\bibfnamefont {F.}~\bibnamefont {Hassler}},\ }\bibfield  {title} {\bibinfo
  {title} {Weak measurement protocols for majorana bound state
  identification},\ }\href {https://doi.org/10.1103/PhysRevLett.124.096801}
  {\bibfield  {journal} {\bibinfo  {journal} {Phys. Rev. Lett.}\ }\textbf
  {\bibinfo {volume} {124}},\ \bibinfo {pages} {096801} (\bibinfo {year}
  {2020})}\BibitemShut {NoStop}%
\bibitem [{\citenamefont {Prada}\ \emph {et~al.}(2017)\citenamefont {Prada},
  \citenamefont {Aguado},\ and\ \citenamefont {San-Jose}}]{twoMBSs7}%
  \BibitemOpen
  \bibfield  {author} {\bibinfo {author} {\bibfnamefont {E.}~\bibnamefont
  {Prada}}, \bibinfo {author} {\bibfnamefont {R.}~\bibnamefont {Aguado}},\ and\
  \bibinfo {author} {\bibfnamefont {P.}~\bibnamefont {San-Jose}},\ }\bibfield
  {title} {\bibinfo {title} {Measuring majorana nonlocality and spin structure
  with a quantum dot},\ }\href {https://doi.org/10.1103/PhysRevB.96.085418}
  {\bibfield  {journal} {\bibinfo  {journal} {Phys. Rev. B}\ }\textbf {\bibinfo
  {volume} {96}},\ \bibinfo {pages} {085418} (\bibinfo {year}
  {2017})}\BibitemShut {NoStop}%
\bibitem [{\citenamefont {Ueda}\ and\ \citenamefont
  {Yokoyama}(2014)}]{twoMBSs8}%
  \BibitemOpen
  \bibfield  {author} {\bibinfo {author} {\bibfnamefont {A.}~\bibnamefont
  {Ueda}}\ and\ \bibinfo {author} {\bibfnamefont {T.}~\bibnamefont
  {Yokoyama}},\ }\bibfield  {title} {\bibinfo {title} {Anomalous interference
  in aharonov-bohm rings with two majorana bound states},\ }\href
  {https://doi.org/10.1103/PhysRevB.90.081405} {\bibfield  {journal} {\bibinfo
  {journal} {Phys. Rev. B}\ }\textbf {\bibinfo {volume} {90}},\ \bibinfo
  {pages} {081405} (\bibinfo {year} {2014})}\BibitemShut {NoStop}%
\bibitem [{\citenamefont {Zeng}\ \emph {et~al.}(2016)\citenamefont {Zeng},
  \citenamefont {Chen}, \citenamefont {You},\ and\ \citenamefont
  {Lü}}]{twoMBSs9}%
  \BibitemOpen
  \bibfield  {author} {\bibinfo {author} {\bibfnamefont {Q.-B.}\ \bibnamefont
  {Zeng}}, \bibinfo {author} {\bibfnamefont {S.}~\bibnamefont {Chen}}, \bibinfo
  {author} {\bibfnamefont {L.}~\bibnamefont {You}},\ and\ \bibinfo {author}
  {\bibfnamefont {R.}~\bibnamefont {Lü}},\ }\bibfield  {title} {\bibinfo
  {title} {Transport through a quantum dot coupled to two majorana bound
  states},\ }\href {https://doi.org/10.1007/s11467-016-0620-3} {\bibfield
  {journal} {\bibinfo  {journal} {Frontiers of Physics}\ }\textbf {\bibinfo
  {volume} {12}},\ \bibinfo {pages} {127302} (\bibinfo {year}
  {2016})}\BibitemShut {NoStop}%
\bibitem [{\citenamefont {Ozaki}(2007)}]{RN203}%
  \BibitemOpen
  \bibfield  {author} {\bibinfo {author} {\bibfnamefont {T.}~\bibnamefont
  {Ozaki}},\ }\bibfield  {title} {\bibinfo {title} {Continued fraction
  representation of the fermi-dirac function for large-scale electronic
  structure calculations},\ }\href {https://doi.org/10.1103/PhysRevB.75.035123}
  {\bibfield  {journal} {\bibinfo  {journal} {Phys. Rev. B}\ }\textbf {\bibinfo
  {volume} {75}},\ \bibinfo {pages} {035123} (\bibinfo {year}
  {2007})}\BibitemShut {NoStop}%
\bibitem [{\citenamefont {Xie}\ \emph {et~al.}(2012)\citenamefont {Xie},
  \citenamefont {Jiang}, \citenamefont {Tian}, \citenamefont {Zheng},
  \citenamefont {Kwok}, \citenamefont {Chen}, \citenamefont {Yam},
  \citenamefont {Yan},\ and\ \citenamefont {Chen}}]{RN204}%
  \BibitemOpen
  \bibfield  {author} {\bibinfo {author} {\bibfnamefont {H.}~\bibnamefont
  {Xie}}, \bibinfo {author} {\bibfnamefont {F.}~\bibnamefont {Jiang}}, \bibinfo
  {author} {\bibfnamefont {H.}~\bibnamefont {Tian}}, \bibinfo {author}
  {\bibfnamefont {X.}~\bibnamefont {Zheng}}, \bibinfo {author} {\bibfnamefont
  {Y.}~\bibnamefont {Kwok}}, \bibinfo {author} {\bibfnamefont {S.}~\bibnamefont
  {Chen}}, \bibinfo {author} {\bibfnamefont {C.}~\bibnamefont {Yam}}, \bibinfo
  {author} {\bibfnamefont {Y.}~\bibnamefont {Yan}},\ and\ \bibinfo {author}
  {\bibfnamefont {G.}~\bibnamefont {Chen}},\ }\bibfield  {title} {\bibinfo
  {title} {{Time-dependent quantum transport: An efficient method based on
  Liouville-von-Neumann equation for single-electron density matrix}},\ }\href
  {https://doi.org/10.1063/1.4737864} {\bibfield  {journal} {\bibinfo
  {journal} {The Journal of Chemical Physics}\ }\textbf {\bibinfo {volume}
  {137}},\ \bibinfo {pages} {044113} (\bibinfo {year} {2012})}\BibitemShut
  {NoStop}%
\bibitem [{\citenamefont {Croy}\ and\ \citenamefont {Saalmann}(2010)}]{RN205}%
  \BibitemOpen
  \bibfield  {author} {\bibinfo {author} {\bibfnamefont {A.}~\bibnamefont
  {Croy}}\ and\ \bibinfo {author} {\bibfnamefont {U.}~\bibnamefont
  {Saalmann}},\ }\bibfield  {title} {\bibinfo {title} {Erratum: Partial
  fraction decomposition of the fermi function [phys. rev. b 80, 073102
  (2009)]},\ }\href {https://doi.org/10.1103/PhysRevB.82.159904} {\bibfield
  {journal} {\bibinfo  {journal} {Phys. Rev. B}\ }\textbf {\bibinfo {volume}
  {82}},\ \bibinfo {pages} {159904} (\bibinfo {year} {2010})}\BibitemShut
  {NoStop}%
\bibitem [{\citenamefont {Hu}\ \emph {et~al.}(2010)\citenamefont {Hu},
  \citenamefont {Xu},\ and\ \citenamefont {Yan}}]{RN119}%
  \BibitemOpen
  \bibfield  {author} {\bibinfo {author} {\bibfnamefont {J.}~\bibnamefont
  {Hu}}, \bibinfo {author} {\bibfnamefont {R.~X.}\ \bibnamefont {Xu}},\ and\
  \bibinfo {author} {\bibfnamefont {Y.}~\bibnamefont {Yan}},\ }\bibfield
  {title} {\bibinfo {title} {Communication: Pade spectrum decomposition of
  fermi function and bose function},\ }\href
  {https://doi.org/10.1063/1.3484491} {\bibfield  {journal} {\bibinfo
  {journal} {J Chem Phys}\ }\textbf {\bibinfo {volume} {133}},\ \bibinfo
  {pages} {101106} (\bibinfo {year} {2010})}\BibitemShut {NoStop}%
\bibitem [{\citenamefont {Karrasch}\ \emph {et~al.}(2010)\citenamefont
  {Karrasch}, \citenamefont {Meden},\ and\ \citenamefont
  {Sch\"onhammer}}]{RN127}%
  \BibitemOpen
  \bibfield  {author} {\bibinfo {author} {\bibfnamefont {C.}~\bibnamefont
  {Karrasch}}, \bibinfo {author} {\bibfnamefont {V.}~\bibnamefont {Meden}},\
  and\ \bibinfo {author} {\bibfnamefont {K.}~\bibnamefont {Sch\"onhammer}},\
  }\bibfield  {title} {\bibinfo {title} {Finite-temperature linear conductance
  from the matsubara green's function without analytic continuation to the real
  axis},\ }\href {https://doi.org/10.1103/PhysRevB.82.125114} {\bibfield
  {journal} {\bibinfo  {journal} {Phys. Rev. B}\ }\textbf {\bibinfo {volume}
  {82}},\ \bibinfo {pages} {125114} (\bibinfo {year} {2010})}\BibitemShut
  {NoStop}%
\bibitem [{\citenamefont {Dong}\ \emph {et~al.}(2015)\citenamefont {Dong},
  \citenamefont {Ding},\ and\ \citenamefont {Lei}}]{RN112}%
  \BibitemOpen
  \bibfield  {author} {\bibinfo {author} {\bibfnamefont {B.}~\bibnamefont
  {Dong}}, \bibinfo {author} {\bibfnamefont {G.~H.}\ \bibnamefont {Ding}},\
  and\ \bibinfo {author} {\bibfnamefont {X.~L.}\ \bibnamefont {Lei}},\
  }\bibfield  {title} {\bibinfo {title} {Time-dependent quantum transport
  through an interacting quantum dot beyond sequential tunneling: second-order
  quantum rate equations},\ }\href
  {https://doi.org/10.1088/0953-8984/27/20/205303} {\bibfield  {journal}
  {\bibinfo  {journal} {J Phys Condens Matter}\ }\textbf {\bibinfo {volume}
  {27}},\ \bibinfo {pages} {205303} (\bibinfo {year} {2015})}\BibitemShut
  {NoStop}%
\bibitem [{\citenamefont {Alomar}\ \emph {et~al.}(2016)\citenamefont {Alomar},
  \citenamefont {Lim},\ and\ \citenamefont {S\'anchez}}]{RC1}%
  \BibitemOpen
  \bibfield  {author} {\bibinfo {author} {\bibfnamefont {M.~I.}\ \bibnamefont
  {Alomar}}, \bibinfo {author} {\bibfnamefont {J.~S.}\ \bibnamefont {Lim}},\
  and\ \bibinfo {author} {\bibfnamefont {D.}~\bibnamefont {S\'anchez}},\
  }\bibfield  {title} {\bibinfo {title} {Coulomb-blockade effect in nonlinear
  mesoscopic capacitors},\ }\href {https://doi.org/10.1103/PhysRevB.94.165425}
  {\bibfield  {journal} {\bibinfo  {journal} {Phys. Rev. B}\ }\textbf {\bibinfo
  {volume} {94}},\ \bibinfo {pages} {165425} (\bibinfo {year}
  {2016})}\BibitemShut {NoStop}%
\end{thebibliography}%

\end{document}